\pdfoutput=1
\documentclass[aps,pre,twocolumn,longbibliography]{revtex4-2}
\usepackage[utf8]{inputenc}
\usepackage{amsmath}
\usepackage{commath}
\usepackage[pdftex]{graphicx}
\usepackage{siunitx}
\usepackage{hyperref}
\usepackage{xr}
\usepackage{makecell}
\usepackage{xcolor}
\usepackage{ulem}

\makeatletter
\newcommand*{\addFileDependency}[1]{% argument=file name and extension
  \typeout{(#1)}
  \@addtofilelist{#1}
  \IfFileExists{#1}{}{\typeout{No file #1.}}
}
\makeatother

\newcommand*{\myexternaldocument}[1]{%
    \externaldocument{#1}%
    \addFileDependency{#1.tex}%
    \addFileDependency{#1.aux}%
}

\myexternaldocument{supplementary_labels}

\usepackage{cleveref}

\begin{document}
\title{Machine learning reveals strain-rate-dependent predictability\\ of discrete dislocation plasticity}
\author{Marcin Mi{\'n}kowski}
\email{marcin.minkowski@tuni.fi}
\author{David Kurunczi-Papp}
\author{Lasse Laurson}
\affiliation{Computational Physics Laboratory, Tampere University, P.O. Box 692, FI-33014 Tampere, Finland}
%\date{January 2021}

\begin{abstract}
Predicting the behaviour of complex systems is one of the main goals of science. An important example is plastic deformation of micron-scale crystals, a process mediated by collective dynamics of dislocations, manifested as broadly distributed strain bursts and significant sample-to-sample variations in the response to applied loading. Here, by combining large-scale discrete dislocation dynamics simulations and machine learning, we study the problem of predicting the fluctuating stress-strain curves of individual small single crystals subject to strain-controlled loading using features of the initial dislocation configurations as input.
Our results reveal an intriguing rate dependence of deformation predictability: For small strains predictability improves with increasing strain rate, while for larger strains the predictability vs strain rate relation becomes non-monotonic. We show that for small strains the rate-dependence of deformation predictability can be captured by considering the fraction of dislocations moving against the direction imposed by the external stress, serving as a measure of strain-rate-dependent complexity of the dislocation dynamics. The non-monotonic predictability vs strain rate relation for large strains is argued to be related to a transition from fluctuating to smooth plastic flow when strain rate is increased.
\end{abstract}

\maketitle

\section{Introduction}

Dislocation-mediated plasticity of crystalline solids is characterized by size effects implying that smaller systems tend to be "stronger" and exhibit larger fluctuations in their stress-strain response~\cite{uchic2009plasticity,dimiduk2005size}, as well as by rate effects which are typically manifested as larger average stresses needed to reach a given strain for higher strain rates~\cite{fan2021strain}, a phenomenon often referred to as strain rate sensitivity~\cite{schwaiger2003some}. Understanding the combined effects of these key features of dislocation plasticity constitutes an important challenge in physics and materials science: For instance, the strain bursts exhibited by micropillars under compression~\cite{dimiduk2006scale} are interesting both from the perspective of fundamentals of non-equilibrium statistical physics~\cite{ispanovity2014avalanches,ovaska2015quenched} as well as in applications where the fluctuations of plastic deformation result in difficulties to control the resulting shape in plastic forming processes~\cite{csikor2007dislocation}. At the same time, the fact that in various applications the mechanical response of metals depends significantly on the loading rate (e.g., during a car crash) is of great importance and calls for novel approaches to understand the underlying fundamental physics and to control rate dependence of dislocation plasticity.

\begin{figure*}[t!]
	\centering
	\includegraphics[width=0.9\textwidth]{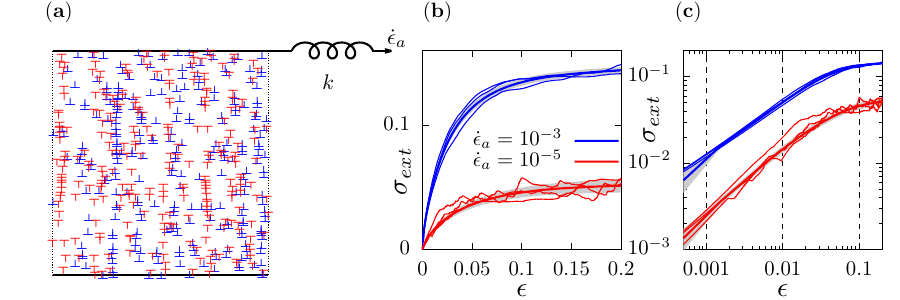}
	\caption{{\bf Schematic of the DDD model with strain-controlled loading, and examples of the resulting stress-strain curves.} (a) Strain-controlled loading of a crystal containing dislocations with positive (red) and negative (blue) Burgers vectors, with the applied strain rate $\dot{\epsilon}_a$ and the stiffness of the specimen-machine system, modelled with a spring of stiffness $k$. (b) Stress-strain curves for a few samples for $k=1$ and two different values of $\dot{\epsilon}_a$ plotted with thin lines in linear scale, with the thick line representing the average over all the samples and the gray area indicating their standard deviation. (c) The same curves in the logarithmic scale, with the vertical dashed lines representing the values of strain for which the predictive algorithms used in the work are trained.}
	\label{scheme_model}
\end{figure*}

Recent years have witnessed a surge in application of artificial intelligence (AI) in general and machine learning (ML) in particular to gain novel insights on properties of materials and related problems in physics~\cite{papanikolaou2019spatial,raissi2018hidden, baldi2014searching, butler2018machine,butler2018machine,pilania2013accelerating,viitanen2020machine,biswas2020prediction,karniadakis2021physics,salmenjoki2018mimicking,salmenjoki2020probing,bock2019review}. Broadly speaking, such developments fall under the umbrella of the emerging research field of {\it materials informatics}~\cite{ramprasad2017machine}, where informatics methods -- including ML -- are used to search for novel materials~\cite{liu2017materials,guo2021artificial}, establish novel structure-property relations~\cite{YU2021102570,pilania2013accelerating}, etc. A closely related problem is given by quantification of {\it deformation predictability}~\cite{salmenjoki2018machine,sarvilahti2020machine} of small solid samples where the response to applied stresses (i.e., the stress-strain curve) exhibits significant sample-to-sample variations. The general problem statement can be formulated as follows: Given a description of the initial microstructure (defined, e.g., by the arrangement of pre-existing dislocations and, if present, other defects within the crystal) of such a sample, how well can its subsequent sample-specific deformation dynamics be predicted when external loading is applied? Such questions are closely related to the more general issue of predictability of "complex systems", ranging from forecasting earthquakes~\cite{rouet2017machine, kagan2000probabilistic,anderson2017theory} to predicting volcanic eruptions~\cite{voight1988method} and beyond~\cite{israeli2006coarse}. 

Such predictions tend to be difficult as the mapping from the various features describing the initial state of the system to its subsequent behavior or dynamics is usually highly complicated. This complexity typically originates from the combination of non-linear nature of the collective dynamics governing the behaviour of the system and the high dimensionality of the feature set characterizing the system's state. On the other hand, while traditional forecasting methods struggle with such complexity, typical ML algorithms are well-suited to address problems of this type. Our recent numerical work based on combining ML and discrete dislocation dynamics (DDD) simulations suggests that under quasistatic stress-controlled loading the degree of deformation predictability of plastically deforming crystals is controlled by the collective nature of dislocation dynamics, manifested as inherently hard to predict dislocation avalanches~\cite{salmenjoki2018machine,sarvilahti2020machine}. It is worth noting that even in situations where the dislocation motion is governed by deterministic equations of motion, the critical-like and possibly chaotic~\cite{deshpande2001dislocation} dynamics exhibited by the system renders the prediction problem highly non-trivial~\cite{sarvilahti2020machine}. These works highlight the usefulness of ML in providing a metric for deformation predictability, allowing also to identify the key features of the initial microstructure that determine the mechanical properties of the sample. Yet, many outstanding questions related to deformation predictability remain to be addressed, including how rate-dependence of plasticity is manifested in predictability of the deformation process.

\begin{figure*}[t!]
	\centering
	\includegraphics[width=0.9\textwidth]{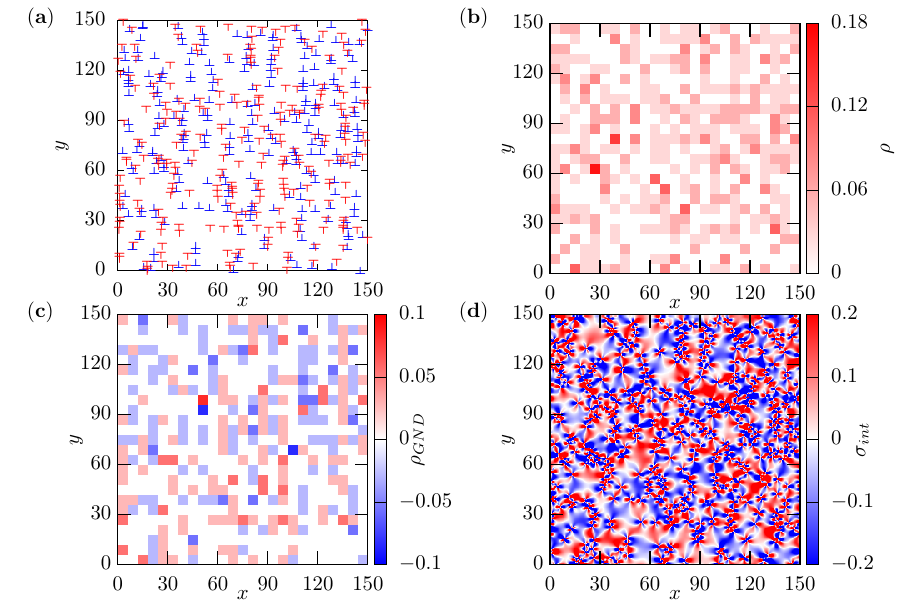}
	\caption{{\bf Fields used as input for linear regression and fully connected neural networks.} (a) An example initial configuration of the dislocations, and (b) the corresponding density field of dislocations $\rho(x,y)$, (c) density of geometrically necessary dislocations $\rho_{GND} (x,y)$, and (d) the internal stress field $\sigma_{int}(x,y)$ produced by the dislocations. Various features derived from these fields (see Methods) are used as input for linear regression and fully connected neural networks.}
	\label{fields}
\end{figure*}

Here, we combine large-scale strain-controlled DDD simulations~\cite{kurunczipapp2021disloc} and ML to study how the imposed strain rate and the stiffness of the specimen-machine system affect deformation predictability of small single crystals. By employing ML models including linear regression, fully connected neural networks and convolutional neural networks, trained using data from our extensive 2D DDD simulations, we find that the ability of all these rather different ML models to predict the stress-strain curve using features of the initial dislocation configuration as input, quantified by the correlation between predicted and actual stress at a given strain, exhibits a similar dependence on the imposed strain rate. Hence, by employing the definition of deformation predictability used in Refs.~\cite{salmenjoki2018machine,sarvilahti2020machine}, we conclude that deformation predictability of the 2D DDD model we study is rate-dependent. For small strains, predictability improves with increasing strain rate, while for large strains the predictability vs strain rate relation becomes non-monotonic. Interestingly, especially for small strains, we demonstrate a clear link between deformation predictability and the mean fraction of dislocations moving against the direction where the applied stress is driving them, such that deformation predictability improves when a smaller fraction of the dislocations move against the direction set by the applied stress. This fraction increases with decreasing strain rate and is a measure of how "interaction-dominated" the dislocation dynamics is, and relates to the sizes and complexity of the multipolar dislocation structures moving collectively during the deformation process. Specifically, our results suggest that the more complex dislocation dynamics for low strain rates results in a larger prediction error of the ML models. We also discuss the effect of varying the stiffness of the specimen-machine system, as well as the role of the various features characterizing the initial dislocation configuration in determining the stress-strain response, including how these give rise to the non-monotonic predictability vs strain rate relation for large strains. The latter feature is also demonstrated to be linked to a transition from fluctuating to smooth stress-strain curves upon increasing the imposed strain rate: At the transition point the stress-strain curve is typically dominated by an individual large intrinsically hard-to-predict dislocation avalanche, resulting in a larger prediction error of the ML models.

The paper is organized as follows: We start by going through the methods employed (Section~\ref{sec:methods}), including the DDD simulations in Subsection~\ref{subsec:DDD}, input features used for the ML models in Subsection~\ref{subsec:features}, and the ML models considered (linear regression, fully connected neural networks and convolutional neural networks in~\ref{subsec:linreg}, \ref{subsec:nn}, and \ref{subsec:cnn}, respectively). This is then followed by presentation of our results in Section~\ref{sec:results}, including details about dataset generation in Subsection~\ref{subsec:dataset}, machine learning results in Subsection~\ref{subsec:ml}, and an account on how these are related to the rate-dependent complexity of dislocation dynamics in Subsections~\ref{subsec:complexity} and~\ref{subsec:nonmonotonic}. The paper is finished with discussion and conclusions in Section~\ref{sec:discussion}.

\section{\label{sec:methods} Methods}

\subsection{\label{subsec:DDD} DDD simulations}

The 2D DDD model we use is similar to the models studied in a number of previous works~\cite{miguel2002dislocation,ispanovity2014avalanches,ovaska2015quenched,kurunczipapp2021disloc}. The dynamics of straight parallel edge dislocation lines oriented along the $z$ direction (represented here by points in the $xy$ plane) and moving within a single slip geometry along the $x$ direction is taken to follow overdamped equations of motion where the dislocation velocities are proportional to the total Peach-Koehler force acting on them, 
\begin{equation}
\frac{\dot{x}_{i}}{Mb}=s_{i}b\left[\sum_{j\neq i}^{N}s_{j}\sigma_{disl}(\mathbf{r}_{i}-\mathbf{r}_{j})+\sigma_{ext}\right],
\label{disl_eq}
\end{equation}
where $N$ is the total number of dislocations in the system, $\mathbf{r}_i=(x_i,y_i)$ is the position of the dislocation $i$, $M$ is the dislocation mobility, $b$ is the length of the Burgers vector and $s_{i}=\pm 1$ is its sign. Each dislocation moves in response to both the externally applied shear stress $\sigma_{ext}$ and the internal shear stress produced by the other dislocations in the system, given by a sum over the other dislocations of $\sigma_{disl}(\mathbf{r})=\sigma_{disl}(x,y)=Dbx(x^2-y^2)/(x^2+y^2)^2$, where $D=\mu/2\pi(1-\nu)$, with $\mu$ and $\nu$ the shear modulus and Poisson ratio, respectively. 
%Eq. (\ref{disl_eq}) is employed in all the DDD simulations used in the work. 
Additionally, during the simulation two dislocations of opposite sign annihilate if the distance between them becomes smaller than $b$, and periodic boundary conditions are implemented. In the simulations all lengths, times and stresses are measured in units of $b$, $(MDb)^{-1}$ and $D$, respectively.

Creating a dataset proceeds by generating configurations containing a certain number of randomly distributed dislocations within a square simulation box, and subsequently relaxing them into a metastable state by running the DDD simulation (i.e., numerically integrating Eq.~(\ref{disl_eq}) for all the dislocations) at zero external shear stress, i.e., $\sigma_{ext}=0$. Because of the aforementioned annihilation processes the number of dislocations in the final state is always smaller than in the unrelaxed configuration. The relaxed state is then used as the initial state for the subsequent strain-controlled loading, during which $\sigma_{ext}$ is applied according to the protocol
\begin{equation}
    \sigma_{ext}=k[\dot{\epsilon}_a t-\epsilon(t)],
    \label{loading_eq}
\end{equation}
where $\epsilon(t)=b/L^2\sum_{i=1}^{N}s_i\Delta x_{i}(t)$ is the strain at time $t$, with $\Delta x_{i}(t)$ denoting the displacement of the $i$th dislocation, $\dot{\epsilon}_a$ is the applied strain rate, and $k$ is the spring stiffness, modelling the combined stiffness of the specimen-machine system~\cite{kurunczipapp2021disloc}. Varying $k$ could mean either that stiffness of the loading machine or the shear modulus of the sample changes, or both of them change simultaneously. Taking the liberty of adjusting $k$ allows us to study its influence on deformation predictability. The above relation means that during the simulation the external stress $\sigma_{ext}$ is adjusted in such a way that the true plastic strain rate $\dot{\epsilon}(t) = b/L^2\sum_{i=1}^{N}s_i\dot{x}_i(t)$ approaches the applied one, $\dot{\epsilon}_a$, in the long-time limit, and thereafter fluctuates in its vicinity. The stiffness $k$ measures how strongly $\sigma_{ext}$ reacts to deviations of $\dot{\epsilon}(t)$ from the applied strain rate $\dot{\epsilon}_a$. Thus, higher values of $k$ help to approach the applied strain rate sooner, and result in $\sigma_{ext}$ exhibiting larger fluctuations. Concurrently, larger $k$-values lead to smaller fluctuations of $\dot{\epsilon}$ around $\dot{\epsilon}_a$. Hence, the finite stiffness $k$ of the specimen-machine system implies that we consider an intermediate case between "hard" (strain-controlled with an infinitely stiff machine) and "soft" (stress-controlled) driving~\cite{kurunczipapp2021disloc}. Fig.~\ref{scheme_model}(a) shows a schematic of the simulation setup, with examples of stress-strain curves obtained from the simulations in Fig.~\ref{scheme_model}(b) and (c) for two different $\dot{\epsilon}_a$-values.

\subsection{\label{subsec:features} Input features for machine learning}

\begin{table}[t!]
    \begin{tabular}{|c|c|}
    \hline
    Field & Features \\ \hline
    $\rho_{GND}$ & $|f_{x1}|$, $|f_{y1}|$, $|f_{xy}|$, kurtosis, skewness \\ \hline
    $|\rho_{GND}|$ & average, kurtosis, skewness \\ \hline
    $\rho$ & $|f_{x1}|$, $|f_{y1}|$, $|f_{xy}|$, average, kurtosis, skewness \\ \hline
    $\sigma_{int}$ & $|f_{x1}|$, $|f_{y1}|$, $|f_{xy}|$, kurtosis, skewness \\ \hline
    $|\sigma_{int}|$ & average, kurtosis, skewness \\ \hline
    \end{tabular}
    \caption{List of all the features used for training linear regression and NN.}
    \label{features_list}
\end{table}

\begin{figure}[t!]
	\centering
	\includegraphics[width=0.9\columnwidth]{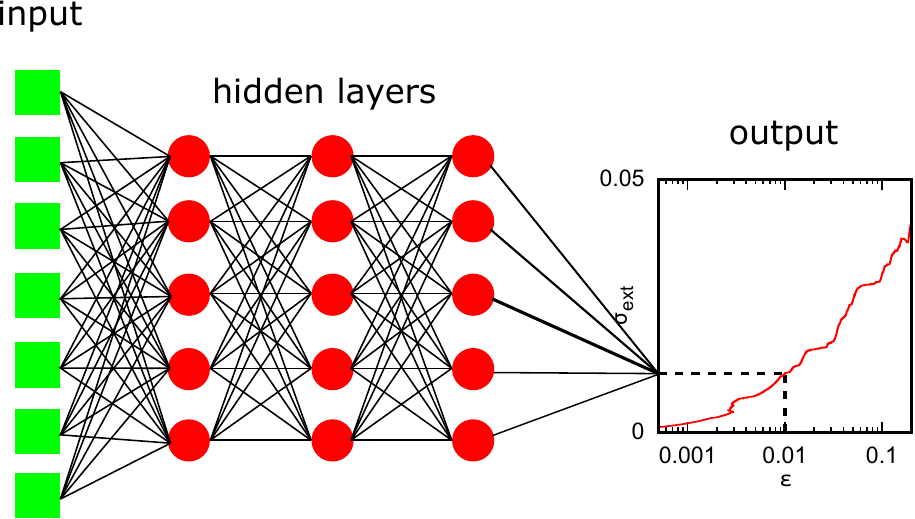}
    \caption{{\bf Schematic representation of the fully connected neural network.} The input values are passed to the neurons in the hidden layers through the activation function to get the external stress value as the final result.}
    \label{NN_scheme}
\end{figure}

\begin{figure*}[t!]
	\centering
	\includegraphics[width=0.8\textwidth]{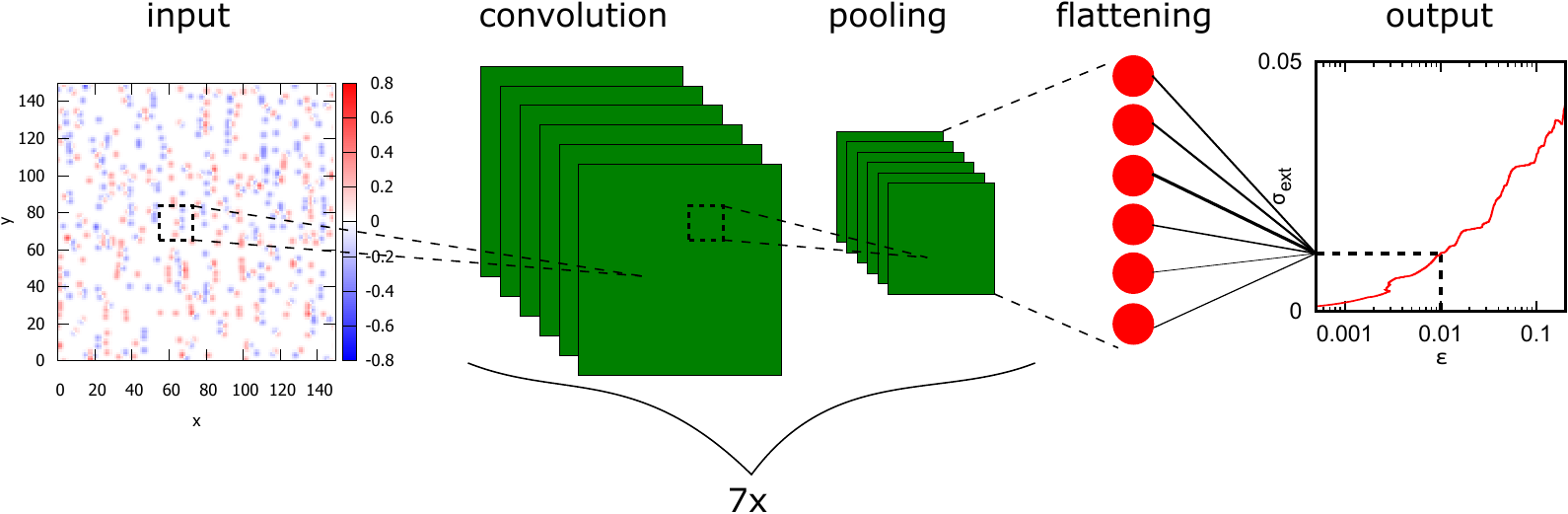}
    \caption{{\bf Schematic representation of the convolutional neural network.} On the left the image representing the dislocation configuration is shown. It is subsequently transformed by the convolutional and the pooling layer. Those two operations are performed 7 times, after which the array is flattened and passed to the fully connected neural network on the right. The final result is the external stress value.}
    \label{CNN_scheme}
\end{figure*}

\begin{table}[t!]
    \begin{tabular}{|l|l|l|}
    \hline
    Layer & \makecell{Shape of \\the output} & \makecell{Number of \\trainable parameters} \\ \hline
    periodic\textunderscore padding\textunderscore layer\textunderscore 1 & 130x130x1 & 0 \\ \hline
    conv2d\textunderscore 1 & 128x128x16 & 160 \\ \hline
    max\textunderscore pooling\textunderscore 1 & 64x64x16 & 0 \\ \hline
    periodic\textunderscore padding\textunderscore layer\textunderscore 2 & 66x66x16 & 0 \\ \hline
    conv2d\textunderscore 2 & 64x64x16 & 2320 \\ \hline
    max\textunderscore pooling\textunderscore 2 & 32x32x16 & 0 \\ \hline
    periodic\textunderscore padding\textunderscore layer\textunderscore 3 & 34x34x16 & 0 \\ \hline
    conv2d\textunderscore 3 & 32x32x16 & 2320 \\ \hline
    max\textunderscore pooling\textunderscore 3 & 16x16x16 & 0 \\ \hline
    periodic\textunderscore padding\textunderscore layer\textunderscore 4 & 18x18x16 & 0 \\ \hline
    conv2d\textunderscore 4 & 16x16x16 & 2320 \\ \hline
    max\textunderscore pooling\textunderscore 4 & 8x8x16 & 0 \\ \hline
    periodic\textunderscore padding\textunderscore layer\textunderscore 5 & 10x10x16 & 0 \\ \hline
    conv2d\textunderscore 5 & 8x8x16 & 2320 \\ \hline
    max\textunderscore pooling\textunderscore 5 & 4x4x16 & 0 \\ \hline
    periodic\textunderscore padding\textunderscore layer\textunderscore 6 & 6x6x16 & 0 \\ \hline
    conv2d\textunderscore 6 & 4x4x16 & 2320 \\ \hline
    max\textunderscore pooling\textunderscore 6 & 2x2x16 & 0 \\ \hline
    periodic\textunderscore padding\textunderscore layer\textunderscore 7 & 4x4x16 & 0 \\ \hline
    conv2d\textunderscore 7 & 2x2x16 & 2320 \\ \hline
    max\textunderscore pooling\textunderscore 7 & 1x1x16 & 0 \\ \hline
    flatten\textunderscore 1 & 16 & 0 \\ \hline
    dense\textunderscore 1 & 1 & 17 \\
    \hline
    \end{tabular}
     \caption{Detailed structure of the CNN with all the layers listed together with the shape of their output and the number of the parameters that are updated during the training.}
    \label{CNN_structure}
\end{table}

In this work three different ML algorithms are used to predict the stress-strain curve based on the initial relaxed dislocation configuration and the parameters of the strain-controlled loading (strain rate $\dot{\epsilon}_a$ and the spring stiffness $k$). These algorithms are linear regression, fully connected neural network (NN) and convolutional neural network (CNN). The first two require manually defined descriptors that characterize the initial state. Those are chosen as the statistical features of fields defined on the configurations: density of dislocations $\rho(x,y)$, density of geometrically necessary dislocations (GND) $\rho_{GND}(x,y)$, defined as the difference between the density of positive and negative dislocations, and the internal stress field due to all the dislocations in the system, $\sigma_{int}(x,y)=\sum_{i}s_{i}\sigma_{disl}(\mathbf{r}-\mathbf{r_i})$. $\rho(x,y)$ and $\rho_{GND}(x,y)$ are sampled over 25 slices in both directions and $\sigma_{int}(x,y)$ over 300 slices. Because the number of positive and negative dislocations in the system is equal, the GND density is obviously zero globally, however, due to the dislocation density fluctuations, it has non-zero values on the local scale. Since the stress field diverges to infinity at the dislocation core, the limit $|\sigma_{int}|<2.0$ is imposed for any $\mathbf{r}$. Fig.~\ref{fields} shows examples of these fields. For the visualisation purpose the limit $|\sigma_{int}|<0.2$ was chosen because most of the values are within this range.

The following features are extracted from the fields: the 1st Fourier coefficient in the $x$ and $y$ directions ($f_{x1}$ and $f_{y1}$), and the mixed coefficient $f_{xy}$ defined for the field $A(x,y)$ as
\begin{eqnarray}
    f_{x1}&=&\sum_{j=0}^{n}\sum_{k=0}^{m}A(j,k)\exp\left[-2\pi i\left(\frac{j}{n+1}\right)\right],\\
    f_{y1}&=&\sum_{j=0}^{n}\sum_{k=0}^{m}A(j,k)\exp\left[-2\pi i\left(\frac{k}{m+1}\right)\right],\\
    f_{xy}&=&\sum_{j=0}^{n}\sum_{k=0}^{m}A(j,k)\exp\left[-2\pi i\left(\frac{j}{n+1}\right)\right]\\
    &&\times\exp\left[-2\pi i\left(\frac{k}{m+1}\right)\right],\nonumber
\end{eqnarray}
where the summation is done over all the slices in both directions, as well as the average, kurtosis and skewness. Since the Fourier coefficients are in general complex numbers, their absolute values are used as the input. The other features are determined both on the original field and its absolute value with the exception of the dislocation density field $\rho(x,y)$, which is non-negative by definition. Moreover, since the averages of the GND density and stress field are by definition equal to 0, only the features defined on the absolute values of those fields are chosen. In total $N_{in}=22$ descriptors are used, with all of them listed in Table~\ref{features_list}.

\subsection{\label{subsec:linreg} Linear regression}

Linear regression is the simplest model employed in this work. It assumes that the external stress is a linear function of all the $N_{in}$ descriptors of the values $x_i$, that is, $\sigma_{ext}=a_{0}+\sum_{i=1}^{N_{in}}a_{i}x_{i}$, where the parameters $a_{i}$ are optimized using the least squares method in such a way that the loss function is as small as possible. The loss function is defined as the sum of the squares of the differences between the predicted and the true external stress value. 80\% of the configurations were used as the training set and 20\% as the test set. Moreover, L2 regularization is utilized, in which a penalty term $\lambda\sum_{i=1}^{N_{in}}a_i^2$, where the factor $\lambda=0.001$, is added to the loss function, however, in this work it does not change the final results.

\subsection{\label{subsec:nn} Fully connected neural network}

Fully connected neural network (NN) employs a series of hidden layers inserted between the input and the output layer. The value of the $m$th node in the $n$th layer, denoted as $y_{m}^{n}$, is in general a non-linear function of values of all the nodes in the $(n-1)$th layer in the following way
\begin{equation}
    y_{m}^{n}=f_{a}^{n-1}(w_{0m}^{n-1}+\sum_{i=1}^{N_{n-1}}w_{im}^{n-1}y_{i}^{n-1}),
\end{equation}
where $f_{a}^{n-1}$ is the activation function between the layers $n-1$ and $n$, $N_{n-1}$ is the number of nodes in the layer $n-1$, and $w_{im}^{n-1}$ are parameters of the NN. For $i\neq 0$ they are called weights and they multiply the values of the nodes in the previous layer, and for $i=0$ they are called biases, which are added to the argument of the activation function.

The schematic representation of the NN used in this work is shown in Fig.~\ref{NN_scheme}. It consists of 3 hidden layers contains 5 neurons each. The rectifier function is chosen as the activation function, except for the output layer, where the linear function is used.

The NN training is performed in Python using the Keras library with the Adam optimizer with the learning rate set to 0.001. During each epoch the parameters of the NN, weights and biases, are updated. 80\% of the configurations constitute the training set, and the rest is divided equally into the test and validation set. The validation set is used as the stopping criterion for the training. The value of the loss function is determined on that set at each epoch and if it increases continuously for over 500 epochs, the training is stopped and the parameters for the lowest loss so far are used. Again, L2 regularization is applied on every layer with $\lambda=0.001$.

\subsection{\label{subsec:cnn} Convolutional neural network}

On the other hand, CNN takes an image of the relaxed configuration as the input, which is later processed by the filters in the convolutional layers. The image is prepared from the field $J$ defined as a sum of Gaussian functions centered at the dislocations \cite{sarvilahti2020machine},
\begin{equation}
    J(\mathbf{r})=\sum_{i=1}^{N}\frac{s_i}{\sqrt{2\pi}}\exp\left(-\frac{|\mathbf{r}-\mathbf{r}_i|^2}{2}\right).
\end{equation}
The architecture of the CNN implemented in this work within the Keras framework as the Sequential model is shown in Fig. \ref{CNN_scheme}. First the input is passed in a form of an array of the size 128$\times$128, corresponding to the chosen resolution of the $J$ field. Then it is converted using a convolution layer with 16 filters of the size 3$\times$3 and stride equal to 1. In order to take into account the periodicity of the system, periodic padding is chosen. Since that padding is not one of the available types of padding in Keras, in this work it is implemented manually. A function which extends the array periodically in all the directions is created and then a separate Keras layer that executes that function is defined and added before each convolutional layer in the CNN. In that way the output of the convolutional layer has the same size as the input. The convoluted array is subsequently passed into a maximum pooling layer with a kernel 2$\times$2 and stride 2, which reduces the size of the array by 2. There are in total 7 sequences of convolutional and pooling layers, after which the size of the array becomes 1$\times$1$\times$16. Finally, after the flattening procedure a fully connected layer with the linear activation function is inserted to obtain a single value as the output. The detailed structure of the CNN with all the layers listed is shown in Table \ref{CNN_structure}. As the previous methods, the CNN training is also performed with L2 regularization with $\lambda=0.001$.

\section{Results}
\label{sec:results}

\subsection{Dataset generation: Strain-controlled DDD simulations}
\label{subsec:dataset}

In order to generate the training data for the machine learning models, we fix the initial number of dislocations to be 900 within a square simulation box of linear size of $L=150$ simulation units, and consider a wide range of driving parameter values, i.e., $k=0.1$, 1 and 10, and $\dot{\epsilon}_a$ varying from $10^{-5}$ to  $10^{-2}$. For all combinations of values of $k$ and $\dot{\epsilon}_{a}$ we study deformation predictability for three values of strain $\epsilon$, i.e., $\epsilon = 0.001$, 0.01, and 0.1. A dataset consisting of 10000 relaxed initial dislocation configurations was created by initializing the systems with 900 randomly positioned dislocations with equal number of dislocations with positive and negative Burgers vectors, which were then allowed to relax in zero external stress to a metastable state. During the initial relaxation and subsequent loading stage a significant fraction of the dislocations get annihilated such that after deformation a typical sample contains roughly half of the initial number of dislocations.

\begin{figure*}[t!]
	\centering
	\includegraphics[width=0.9\textwidth]{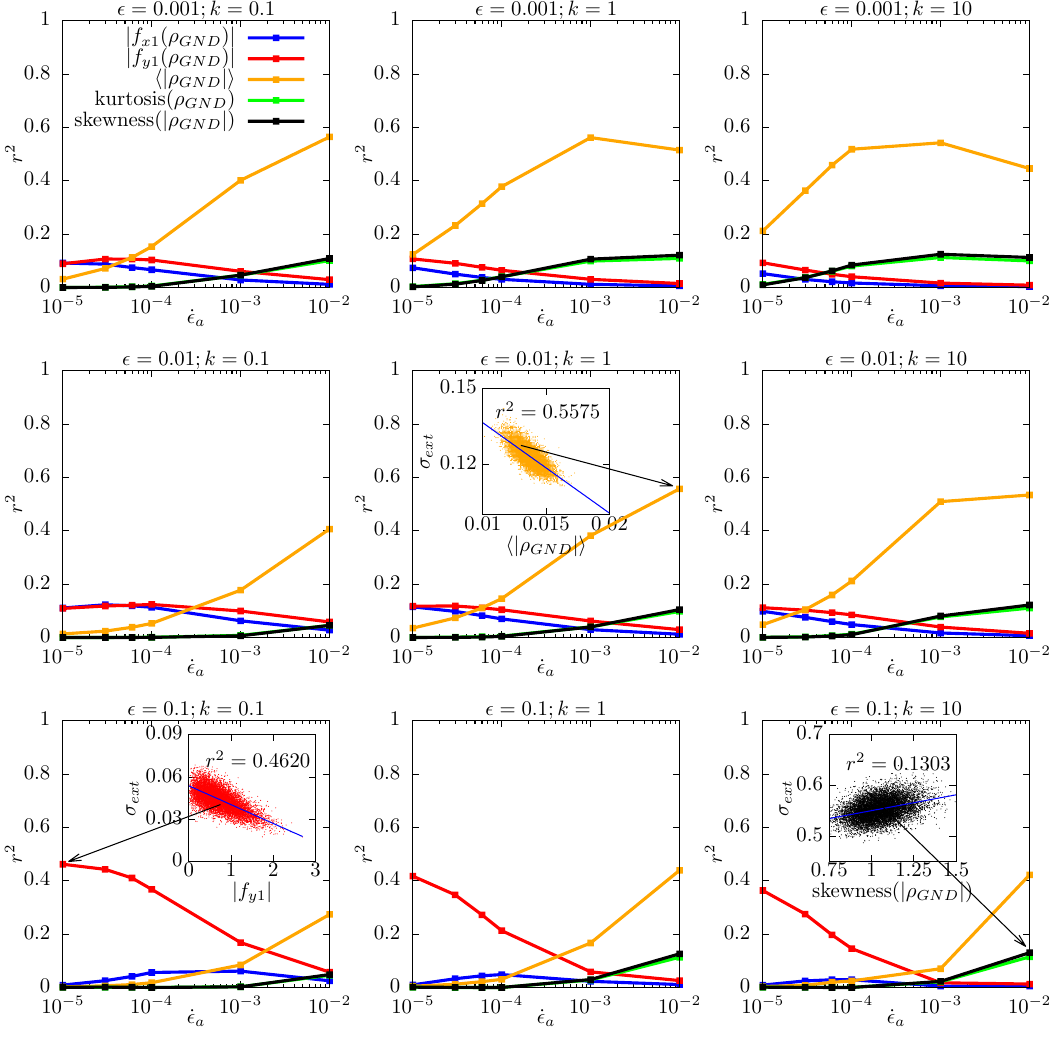}
	\caption{{\bf Correlation of $\rho_{GND}(x,y)$ with the stress-strain response.} Coefficients of linear correlation $r^2$ between various properties of the geometrically necessary dislocations field $\rho_{GND}(x,y)$ and the values of the external stress $\sigma_{ext}$ obtained at given strain $\epsilon$, stiffness $k$ and applied strain rate $\epsilon$. Only the features exhibiting significant correlation are shown. The insets show example fits of the chosen properties at the points indicated by the arrow. Corresponding figures showing the $r^2$'s for the dislocation density field $\rho(x,y)$ and internal stress field $\sigma_{int}$ are included as Supplementary Figs.~\ref{linear_correlations_density} and \ref{linear_correlations_stress_field} \cite{SM}, respectively.}
	\label{linear_correlations_GND}
\end{figure*}

\begin{figure*}[t!]
	\centering
	\includegraphics[width=0.9\textwidth]{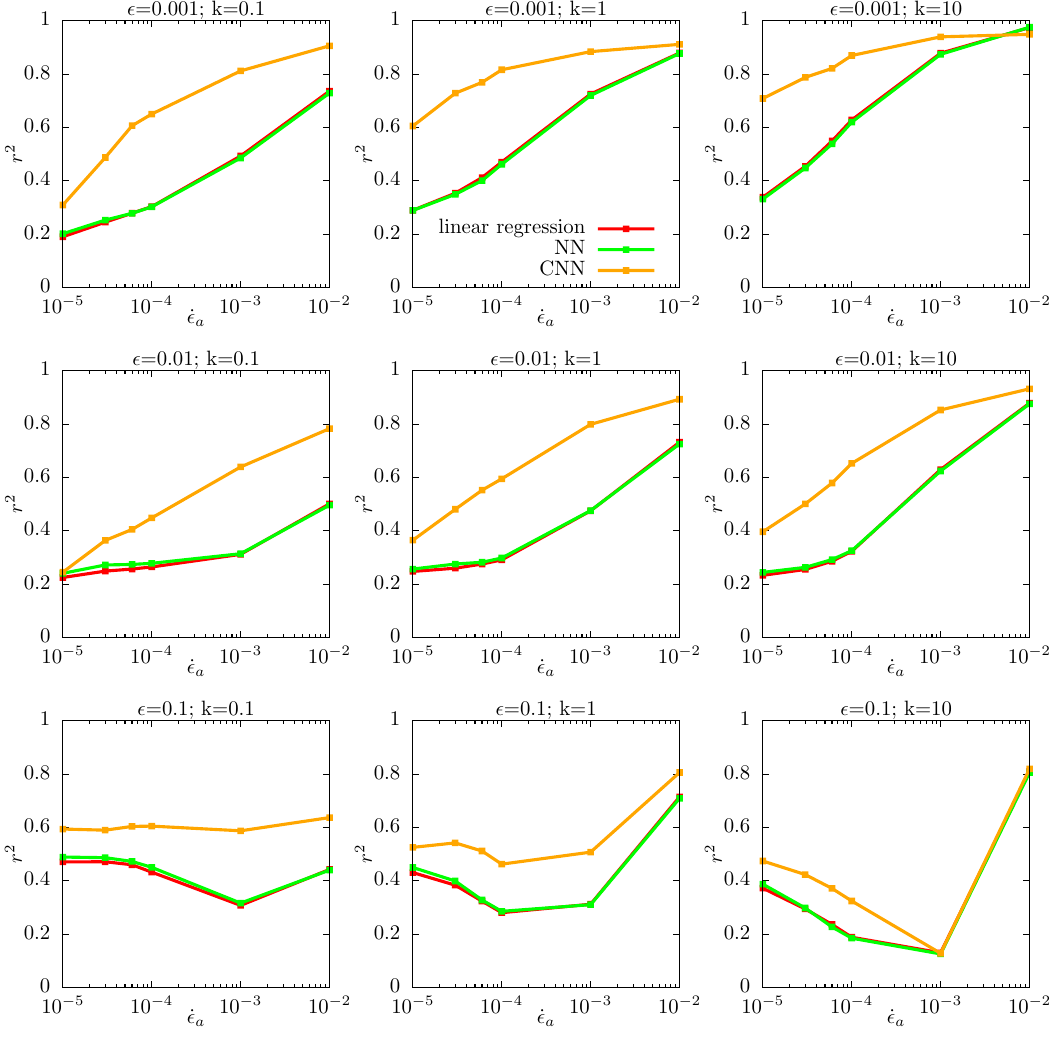}
    \caption{{\bf Rate-dependent deformation predictability revealed by machine learning.} Comparison of test set predictability scores averaged over 5 random seeds for the three different ML methods considered. Notice how predictability improves with increasing $\dot{\epsilon}_a$ for small $\epsilon$, and exhibits non-monotonic dependence on $\dot{\epsilon}_a$ for large $\epsilon$. Note also how linear regression and NN results are almost indistinguishable, while CNN generally outperforms the other two predictive models.}
    \label{test_set_scores}
\end{figure*}

Stress-strain curves were then generated from all of these initial configurations for each combination of the driving parameters $\dot{\epsilon}_a$ and $k$ by simulating the DDD model. See Fig.~\ref{scheme_model} for an example of a dislocation configuration, and ensemble-averaged as well as individual stress-strain curves corresponding to different initial states and strain rates. One may note that each sample has its own, unique fluctuating stress-strain curve, consisting of a sequence of stress drops (due to the instantaneous strain rate $\dot{\epsilon}$ exceeding the externally applied strain rate $\dot{\epsilon}_a$) separated by segments where $\sigma_{ext}$ increases which happens when $\dot{\epsilon}<\dot{\epsilon}_a$ [see Fig.~\ref{scheme_model}(b)]. At the beginning of the deformation the dislocations are still very close to their local minima in the relaxed configuration, therefore the harmonic approximation is valid, which results in the linear dependence of $\sigma_{ext}$ on $\epsilon$. The system is then deformed elastically and returns to the initial state if the stress is removed. For slightly higher strains the system enters the plastic regime, which is characterised by the collective dislocation dynamics. A key feature of that regime is that for small strains the average of the applied stress $\sigma_{ext}$ exhibits a power-law dependence on $\epsilon$ [Fig.~\ref{scheme_model}(c)]; Supplementary Fig.~\ref{power_fits} \cite{SM} shows that the related power-law exponent depends on $k$ and $\dot{\epsilon}_{a}$. Notice also the clear rate effect in the stress-strain curves (Fig.~\ref{scheme_model}): higher $\dot{\epsilon}_{a}$ implies a larger average stress $\sigma_{ext}$ at a given strain $\epsilon$~\cite{kurunczipapp2021disloc}. The resulting dataset, consisting of the relaxed initial dislocation configurations and the corresponding stress-strain curves for different values of $\dot{\epsilon}_a$ and $k$, was then used to train the ML algorithms to infer the link between the features of the initial relaxed dislocation configurations and the stress-strain curves for all combinations of the $\dot{\epsilon}_{a}$ and $k$ values considered.

\subsection{Machine learning rate-dependent dislocation dynamics}
\label{subsec:ml}

Predicting the shape of the stress-strain curve by means of ML algorithms with and without user-defined features allows us to assess how much the results depend on the choice of a specific ML model, and if it is possible to improve the predictions by extending the feature set beyond the obvious one motivated by physical intuition. As described in Sec. \ref{subsec:features}, in this work those features were extracted from three fields defined on a configuration. They should in principle provide a full description of the system, however, in practice we consider mainly large-scale statistical quantities derived from these fields, leaving some room for improvement if local details and correlations turn out to be important. The coefficients of the linear correlation between the selected features of the fields and the corresponding values of the external stress $\sigma_{ext}$ at a given strain $\epsilon$ are defined as
\begin{equation}
    r^2=1-\frac{\sum_{i}[\sigma_{ext}(i)-\sigma_{ext}^{fit}(i)]^2}{\sum_{i}[\sigma_{ext}(i)-\langle \sigma_{ext}(i)\rangle]^2},\label{linear_corr_coeff}
\end{equation}
where the sums are performed over all the configurations and $\sigma_{ext}^{fit}(i)$ is the value of the fitted linear function of the given feature. They are shown in Fig.~\ref{linear_correlations_GND} for the GND density field and in Supplementary Figs.~\ref{linear_correlations_density} and \ref{linear_correlations_stress_field} \cite{SM} for the other two fields. Depending on the driving parameters ($\dot{\epsilon}_a$ and $k$) and the value of $\epsilon$ considered, different features were found to be important. Generally, at small $\epsilon$ and high $\dot{\epsilon}_a$ the average of the absolute value of the GND density, $\langle |\rho_{GND}|\rangle$, correlates strongly with $\sigma_{ext}$, while at large $\epsilon$ and low $\dot{\epsilon}_a$ the 1st Fourier coefficient of $\rho_{GND}$ in the $y$-direction (i.e., $f_{y1}(\rho_{GND})$, see also Ref.~\cite{salmenjoki2018machine}) is more important. As seen in the insets, both of those features correlate negatively with $\sigma_{ext}$, however, for other features, for instance the skewness of $|\rho_{GND}|$, the correlation can be positive. 

In contrast to linear regression and NN, CNN does not require any manually defined features. Instead, a pixelated image of the dislocation configuration is passed as the input. In principle, all the relevant features to the given problem can be extracted from those images by the CNN without user input, and hence CNN has potential to do better at predicting the deformation dynamics than the other two models using user-defined features as input. Here images of the resolution of $128 \times 128$ were used as input for the CNN (see an example input image in Fig.~\ref{CNN_scheme}), chosen in order to be able to resolve individual dislocations while at the same time minimizing the size of the image.

The ML models were trained with 5 different random seeds, which determine how the dataset is split into the training, test and validation set at their fixed ratio of 80:10:10\% (80:20:0\% for linear regression). Our main result, i.e., the predictability scores defined analogously to the linear correlation coefficient, Eq. (\ref{linear_corr_coeff}), determined for the test set of all the ML models and averaged over all the seeds are plotted in Fig.~\ref{test_set_scores}. Scores for all sets (training, test and validation) are shown in Supplementary Fig.~\ref{LR_scores_10000} for linear regression, and in Supplementary Fig.~\ref{NN_scores_10000} \cite{SM} for NN. For CNN, we consider also the dependence of the results on the dataset size: The CNN scores for different sizes of the dataset are shown in Supplementary Figs.~\labelcref{CNN_scores_5000,CNN_scores_6000,CNN_scores_7000,CNN_scores_8000,CNN_scores_9000,CNN_scores_10000}, with the average gap between the training and test set as a function of the dataset size shown in Supplementary Fig.~\ref{CNN_gap} \cite{SM}. These results show that the training of linear regression and NN have converged well for the size of the dataset at hand (10000 samples) in that the scores for the training and test sets are almost equal (no overfitting). On the other hand, CNN could use more data to fully close the gap between the training and test set: linear extrapolation in Supplementary Fig.~\ref{CNN_gap} \cite{SM} suggests that a dataset size of approximately 17000 would be needed for that.

The scores for linear regression and NN are always very close to each other, even though the latter method allows for non-linear relations between the input and output data, suggesting that their performance reflects the set of input features provided rather than differences in the ML models. On the other hand, even if its training did not fully converge using the available dataset, CNN almost always outperforms the other two methods, indicating that it is able to extract additional relevant features not included in the manually engineered features used for linear regression and NN. Hence, in what follows, we take the CNN score to be the best available estimate of how well the stress-strain curves can be predicted using information of the initial dislocation configurations as input. Notably, in most cases the resulting deformation predictability increases with the strain rate, especially at lower strains. However, at the highest strain studied, i.e., for $\epsilon=0.1$, this relation becomes non-monotonic. Especially for the largest $k$ considered, the predictability score $r^2$ starts at a relatively high value followed by a decrease, reaching the minimum at an intermediate strain rate, and finally $r^2$ increases again for the highest strain rate considered. It is worth noting that for small strains and high strain rates, deformation predictability is almost perfect ($r^2$ close to 1), and the large-rate $r^2$ exceeds 0.8 even for the largest $\epsilon$ of 0.1 largely independently of the ML model. Decreasing $\dot{\epsilon}_a$ results in general in a significant reduction of $r^2$ and CNN outperforming the other predictive models. These observations are generally consistent with the values of linear correlation coefficients for the selected features (see Fig.~\ref{linear_correlations_GND} and Supplementary Figs.~\ref{linear_correlations_density} and \ref{linear_correlations_stress_field}) \cite{SM}: for most of those features the correlation increases with the strain rate. The ML models, combining information from all of these features into a single prediction, are in most cases able to do significantly better than just considering the correlation of $\sigma_{ext}(\epsilon)$ with the individual features.

\begin{figure*}[t!]
	\centering
	\includegraphics[width=0.8\textwidth]{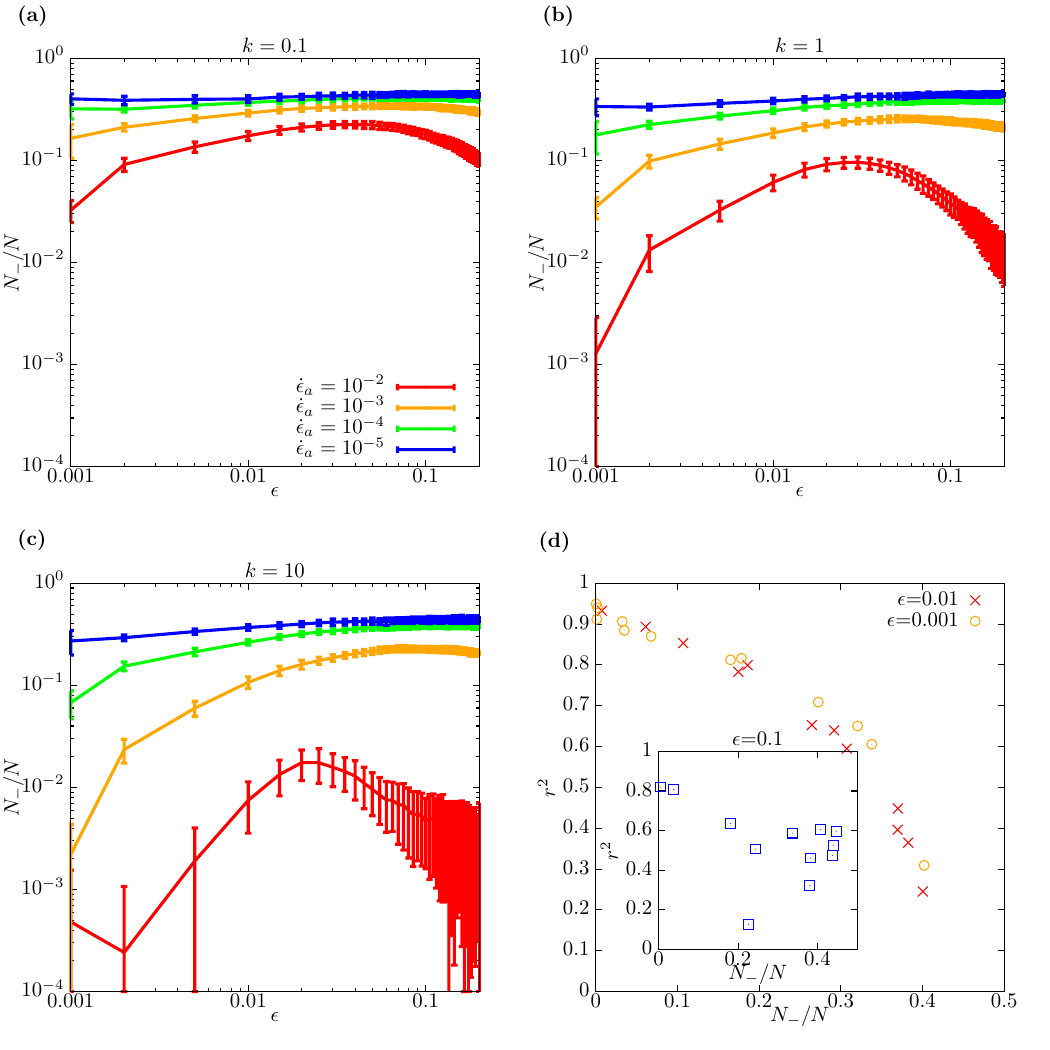}
    \caption{{\bf Rate-dependent complexity of dislocation dynamics controls deformation predictability.} Fraction of dislocations moving in the direction opposite to the one set by the applied external stress, $N_{-}/N$, as a function of $\epsilon$ for (a) $k=0.1$, (b) $k=1$, and (c) $k=10$, averaged over 100 configurations. (d) Shows the correlation between the $r^2$ of the CNN test set and $N_{-}/N$ at the values of $\epsilon$ equal to 0.001, 0.01, and additionally 0.1 in the inset. Error bars are standard error of the mean (SEM).}
    \label{opposite_direction_flow}
\end{figure*}

\begin{figure*}[t!]
	\centering
	\includegraphics[width=0.8\textwidth]{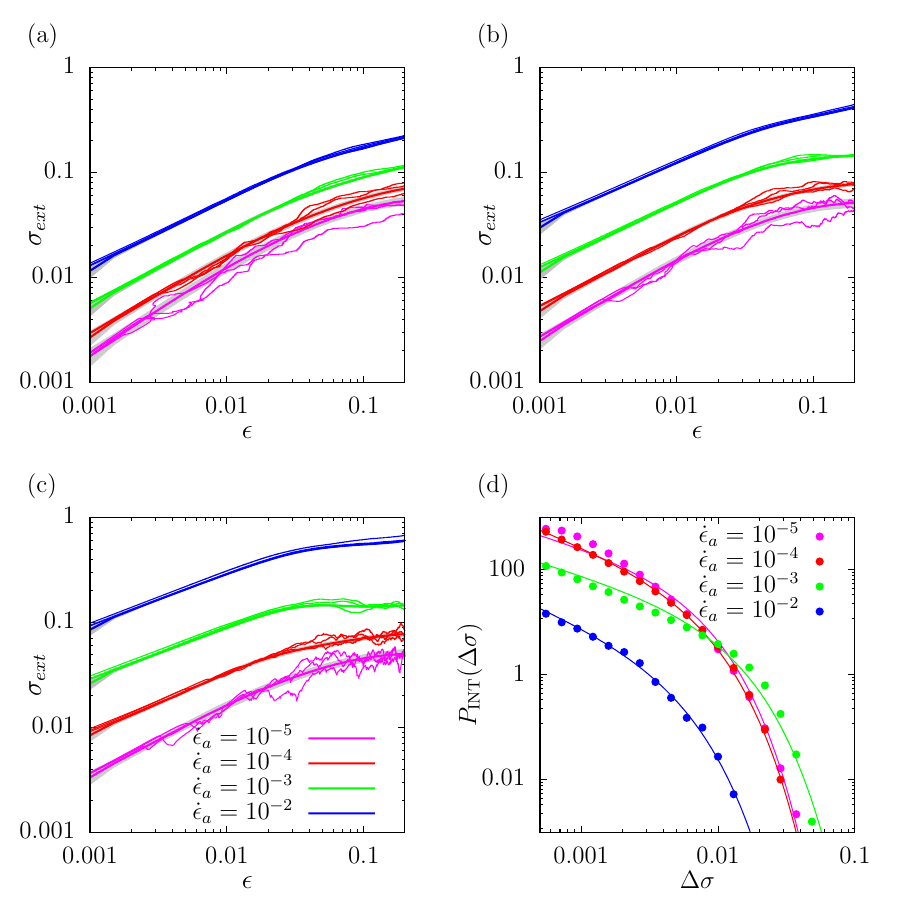}
    \caption{{\bf Stress-strain curves and distribution of stress drops as a function of their magnitude.} (a) $k=0.1$, (b) $k=1$ and (c) $k=10$. (d) Shows the integrated over the whole stress-strain curves of all the samples distribution of stress drops during the loading as a function of their magnitude for $k=10$. The curves represent the power law distribution with an exponential cutoff as expressed by Eq. (\ref{power_law_distribution}) fitted to the data.}
    \label{stress_strain_drops}
\end{figure*}

\subsection{Rate-dependent complexity of dislocation dynamics controls deformation predictability for small strains}
\label{subsec:complexity}

The evolution of deformation predictability with $\dot{\epsilon}_a$ should be related to changes in the fundamental nature of dislocation dynamics as the strain rate is changed. One can expect that at higher strain rates the dynamics becomes simpler -- and hence easier to predict -- because it is dominated by the strong external stress, due to the rate effect discussed above implying higher average $\sigma_{ext}$ at a given $\epsilon$ as $\dot{\epsilon}_a$ is increased. Therefore the internal stresses $\sigma_{int}$ due to the other dislocations are in relative terms less important compared to $\sigma_{ext}$ for high $\dot{\epsilon}_a$. This implies that for high $\dot{\epsilon}_a$ most dislocations move in the direction set by $\sigma_{ext}$ largely independently of the other dislocations. On the other hand, for low $\dot{\epsilon}_a$, $\sigma_{ext}$ is typically smaller and dislocation-dislocation interaction stresses are in relative terms more important, implying complex (and hence likely harder-to-predict) dislocation dynamics where large multipolar dislocation structures slowly drift in the direction set by their net Burgers vector~\cite{miguel2002dislocation}. Hence, for low $\dot{\epsilon}_a$, significant fraction of the dislocations, being part of such collectively moving complex multipolar dislocation structures, are expected to move against the direction set by $\sigma_{ext}$, with strain accumulating due to only a small majority of the dislocations moving in the direction set by external loading. To quantify the rate-dependence of this effect, we consider in Figs.~\ref{opposite_direction_flow}(a), (b) and (c) the ratio of the dislocations moving against the external stress $N_{-}$ and the total number of dislocations $N$ as a function of $\dot{\epsilon}_a$ for different values of $k$.
All the curves start from zero at $\epsilon=0$ (linear response of dislocations in a metastable configuration to a small $\sigma_{ext}$), and plateau at a certain value that is higher for lower $\dot{\epsilon}_a$, implying gradual build-up of dislocation structures of increasing size moving collectively in a direction set by their net Burgers vector. The exception is the curve for the highest strain rate $\dot{\epsilon}_{a}=0.01$ considered, which after reaching a maximum starts to drop significantly, implying that the collectively moving dislocation structures formed during the early stages of loading break up as $\sigma_{ext}$ increases with increasing $\epsilon$. The effect becomes stronger for larger values of $k$, for which the system is stiffer and the strain rate is more accurately controlled by larger fluctuations of $\sigma_{ext}$ as indicated by Eq. (\ref{loading_eq}). As a consequence, for higher $k$ fewer dislocations move against the external stress, especially for larger values of $\dot{\epsilon}_{a}$.

Remarkably, Fig.~\ref{opposite_direction_flow}(d) shows that the CNN predictability score $r^2$ exhibits a very clear negative correlation with $N_{-}/N$ for low strains. Notice how the data points corresponding to different values of $\dot{\epsilon}_a$ and $k$ fall on the same curve independently of whether $\epsilon=0.001$ or 0.01 is considered. This supports the idea that dislocation dynamics becomes more complex and hence harder to predict when smaller $\dot{\epsilon}_a$ and $k$-values are considered. This correlation is weaker for the largest strain value considered, i.e., $\epsilon=0.1$ [inset of Fig.~\ref{opposite_direction_flow}(d)], and hence we need to consider other measures of complexity of the dislocation dynamics in that case.

\subsection{Non-monotonic predictability for large strains
\label{subsec:nonmonotonic}}

In order to account for the non-monotonic dependence of $r^2$ on $\dot{\epsilon}_a$ for large $\epsilon$, we start by considering
the role of different features to understand the origin of the non-monotonic evolution of deformation predictability with $\dot{\epsilon}_a$ in that case.
As seen in Fig.~\ref{linear_correlations_GND}, the 1st Fourier coefficient of $\rho_{GND}$ in the $y$-direction, $f_{y1}(\rho_{GND})$, exhibits for $\epsilon=0.1$ a large $r^2$ for small $\dot{\epsilon}_a$, which then decreases towards zero with increasing $\dot{\epsilon}_a$. $f_{y1}(\rho_{GND})$ is conserved during the dynamics, and hence it is natural that it becomes relevant for large $\epsilon$~\cite{salmenjoki2018machine}. At the same time, for many other features [notably for $\langle |\rho_{GND}| \rangle$ (see Fig.~\ref{linear_correlations_GND}), $\langle \rho \rangle$ (Supplementary Fig.~\ref{linear_correlations_density}), and several features of the $\sigma_{int}$ field (Supplementary Fig.~\ref{linear_correlations_stress_field})] \cite{SM}, $r^2$ increases with $\dot{\epsilon}_a$, such that they become important when $f_{y1}(\rho_{GND})$ ceases to be relevant for large $\dot{\epsilon}_a$, naturally resulting in non-monotonic dependence of the overall $r^2$ on $\dot{\epsilon}_a$ for the ML models using these features as input.

In order to understand the non-monotonic dependence of $r^2$ on $\dot{\epsilon}_a$ at high $\epsilon$ more qualitatively one can analyse the evolution of the corresponding stress-strain curves, keeping in mind that individual critical-like dislocation avalanches are typically understood to be inherently hard-to-predict~\cite{sarvilahti2020machine}. In Figs.~\ref{stress_strain_drops}(a), (b) and (c) example stress-strain curves are shown for $k=$ 0.1, 1, and 10, respectively, for the four largest strain rates studied. Generally, the curves become more regular for larger values of $\dot{\epsilon}_a$. Those for $\dot{\epsilon}_a=10^{-2}$ contain no significant stress drops, therefore, they are the easiest to predict by the ML algorithm since the stress increases monotonically with the strain without any hard-to-predict stress drops. On the other hand, the stress-strain curves for the lowest strain rates tend to be very irregular and exhibit a large number of stress drops. In such cases predicting the stress value requires estimating the number and magnitude of those events occurring prior to the given strain value. However, it is easier to perform when those events are more frequent because their occurrences are then closer to some averaged distribution described by a certain function. It is also expected that recent observations of correlations between subsequent avalanche events driving the individual curves towards the mean ones~\cite{PhysRevMaterials.5.073601} might play a role here. In Fig.~\ref{stress_strain_drops}(d) the distributions of stress drops as a function of their magnitude are shown for different $\dot{\epsilon}_a$ in the case of $k=10$, i.e., the $k$-value for which the non-monotonic behaviour of $r^2$ is most pronounced. The fitted curves (shown as lines) represent the power law distribution with an exponential cutoff~\cite{kurunczipapp2021disloc},
\begin{equation}
    P_{\text{INT}}(\Delta\sigma)=a(\Delta\sigma)^{-\tau_{\sigma,\text{INT}}}\exp(-\Delta\sigma/\Delta\sigma_{0}),\label{power_law_distribution}
\end{equation}
where $a$, $\tau_{\sigma,\text{INT}}$ and $\Delta\sigma_{0}$ are fitting parameters. The largest drops occur for $\dot{\epsilon}_a=10^{-3}$, for which $r^2$ has the minimum. However, there are also fewer small drops than for $\dot{\epsilon}_a=10^{-4}$. Generally, one can expect that strain drops that occur rarely, that is, much less often than once per configuration, do not have a significant detrimental effect on the predictability. On the other hand, the number of events that occur very frequently can also be predicted relatively easily because the estimation error becomes smaller. The hardest to predict are stress-strain curves with events that occur on average once per stress-strain curve: For accurate estimation of the stress at a given strain which is in that case largely controlled by the occurrence of an individual stress drop, the algorithm must predict whether the event will occur before or after a certain value of strain, and what the magnitude of that event will be. Since for $\dot{\epsilon}_a=10^{-3}$ the stress drops are rare, and as seen in Fig.~\ref{stress_strain_drops}(d), their distribution has a larger cutoff than for the other strain rates, the corresponding stress-strain curves are also the most difficult to predict. Interestingly, the minimum of $r^2$ (or, the maximum of the prediction error) thus appears to coincide with a transition from fluctuating to smooth plastic flow upon increasing $\dot{\epsilon}_a$.\\

\section{Discussion and conclusions}
\label{sec:discussion}

To conclude, we have established that predictability of strain-controlled deformation of plastically deforming single crystals is rate-dependent. %observed a strain-rate effect on the deformation predictability in strain-rate controlled loading. 
While the predictability scores of CNN, using images of the configurations as input without explicit feature engineering by the user, are generally higher than those of linear regression and NN using manually engineered input features, all the algorithms exhibit very similar trends of $r^2$ as a function of $\dot{\epsilon}_a$, $\epsilon$, and $k$: Better deformation predictability (i.e., lower prediction error of the ML models) with increasing $\dot{\epsilon}_a$ at low $\epsilon$, and non-monotonic relation between predictability and $\dot{\epsilon}_a$ for high $\epsilon$, especially for large driving spring stiffness $k$. These results suggest that deformation predictability as measured by the ML models is related at least partially to some intrinsic aspects of the dislocation system and the deformation process studied. Indeed, we showed that at lower strains a linear correlation between the predictability and the fraction of dislocations moving against the external stress, which we chose as a measure of the complexity of the deformation process, exists. Since for high $\dot{\epsilon}_a$ $\sigma_{ext}$ is also high, most dislocations move in the direction of the applied stress, and the dislocation dynamics is "simple", and as our results suggest, easier to predict. On the other hand, for lower strain rates the dislocations are more free to follow the system's internal dynamics, governed mostly by the interactions between the dislocations. Since $f_{y1}(\rho_{GND})$ does not change during the dynamics and hence encodes some constant property of the large-scale distribution of the dislocations, it may be relevant for describing those interactions. Indeed, we found that $f_{y1}(\rho_{GND})$ is strongly correlated with $\sigma_{ext}(\epsilon)$ for large $\epsilon$ and small $\dot{\epsilon}_a$, something that together with the correlations exhibited by the other features (which generally increase with increasing $\dot{\epsilon}_a$) allows to understand the non-monotonic dependence of deformation predictability on $\dot{\epsilon}_a$ for large $\epsilon$. On the other hand, this non-monotonic dependence of $r^2$ on $\dot{\epsilon}_a$ for large $\epsilon$ was also argued to be linked to a transition from fluctuating to smooth plastic flow, such that at the transition point the large-strain part of the stress-strain curve is largely controlled by a single hard-to-predict stress drop. In addition to deformation predictability, the degree of complexity of the deformation process evolving with the strain rate outlined above could have more general implications in understanding rate effects in plasticity.

We emphasize that even if the DDD model we have studied obeys deterministic equations of motion, the complex and possibly chaotic nature of the dislocation dynamics implies that perfect predictability (i.e., $r^2=1$) is not reachable in practice, i.e., the prediction error of any real ML model remains finite. One could of course in theory envisage a hypothetical ML model that would be exactly equivalent to a numerical integration algorithm of the equations of motion of the dislocations and would use the initial dislocation positions given with infinite precision as input, and obviously such a model would result in $r^2=1$. However, as soon as there is any deviation from perfect accuracy, e.g., in the initial positions of the dislocations, or if the model does not know the equations of motion perfectly, the perfect predictability is expected to break down. When considering the prospect of studying deformation predictability experimentally, e.g., by extracting descriptors of the initial dislocation microstructure by means of X-ray measurement techniques~\cite{ludwig2001three,levine2006x,groma2013asymmetric}, the interesting question is to what extent a characterization of the initial state with a finite accuracy is suitable for making predictions of the subsequent deformation dynamics. Our results suggest that the answer to this question depends on the strain rate, and it would be interesting to verify this prediction experimentally. 

These results could be generalized to three-dimensional systems with flexible, curved dislocation lines with arbitrary Burgers vectors~\cite{lehtinen2016glassy}. While the dynamics would be more complex and dislocations would have more degrees of freedom, one can expect that the same effect of predictability increasing with strain rate could be observed. That is because (as in the 2D system considered here) the complexity of the dislocation dynamics would be expected to decrease at higher strain rates such that the dislocations are increasingly forced by the external stress to move in a given direction. On the other hand, it would be interesting to study what happens at low strain rates in such systems and if the non-monotonic relation between strain rate and predictability at high strain still exists. While $f_{y1}(\rho_{GND})$ would no longer be conserved during the dynamics in a multi-slip system, there may exist other features that would control the dynamics at low strain rates and large strains. In particular, one could consider dynamics in the presence of pinning centres~\cite{salmenjoki2020plastic}, whose distribution would be an intrinsic feature of the system, which may increase the predictability at larger strains~\cite{sarvilahti2020machine}. 
The above-described extension to 3D DDD would of course be computationally quite demanding, due to increased computational cost of generating the training data set as well as increased complexity of the required ML models. As dislocations would be able to form more complex structures, in the ML models one would need to incorporate additional features related to existence of dislocation junctions, locks, etc. Analogously to the two-dimensional model, one could prepare a three-dimensional voxelized representation of the dislocation configuration to train a CNN model. Another natural extension of the study would be to consider polycrystals containing several crystal grains. This would imply incorporating additional fundamental features to the ML models such as crystal structure, presence of grain boundaries and point defects, etc. Finally, it would also be of interest to consider possible rate effects in the predictability of deformation of amorphous solids~\cite{karmakar2010predicting,richard2020predicting}.

\begin{acknowledgments}
The authors acknowledge the support of the Academy of Finland via the Academy Project COPLAST (project no. 322405). 
\end{acknowledgments}

\bibliography{manuscript}

\end{document}

\author{Marcin Mi{\'n}kowski, David Kurunczi-Papp, Lasse Laurson}
\title{Supplementary material for "Machine learning reveals strain-rate-dependent predictability of discrete dislocation plasticity"}
\maketitle
\begin{figure}[ht]
	\centering
	\resizebox{0.95\columnwidth}{!}{\includegraphics{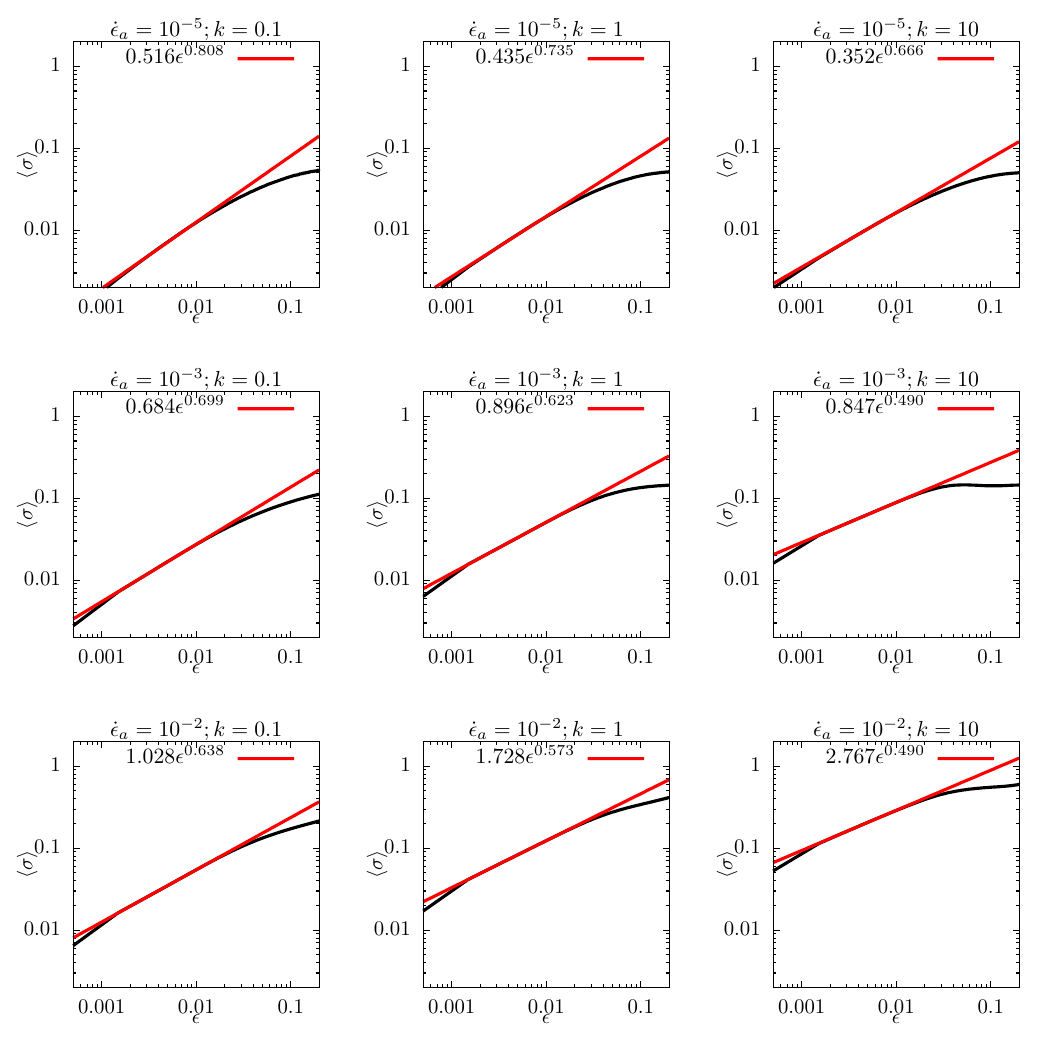}}
	\caption{Power function fits to the averaged stress-strain curves generated at various values of stiffness and strain rate. The fitting was performed in the range of the strain $\epsilon$ equal to (0.001;0.01).}
	\label{power_fits}
\end{figure}
\begin{figure}[ht]
	\centering
	\resizebox{0.95\columnwidth}{!}{\includegraphics{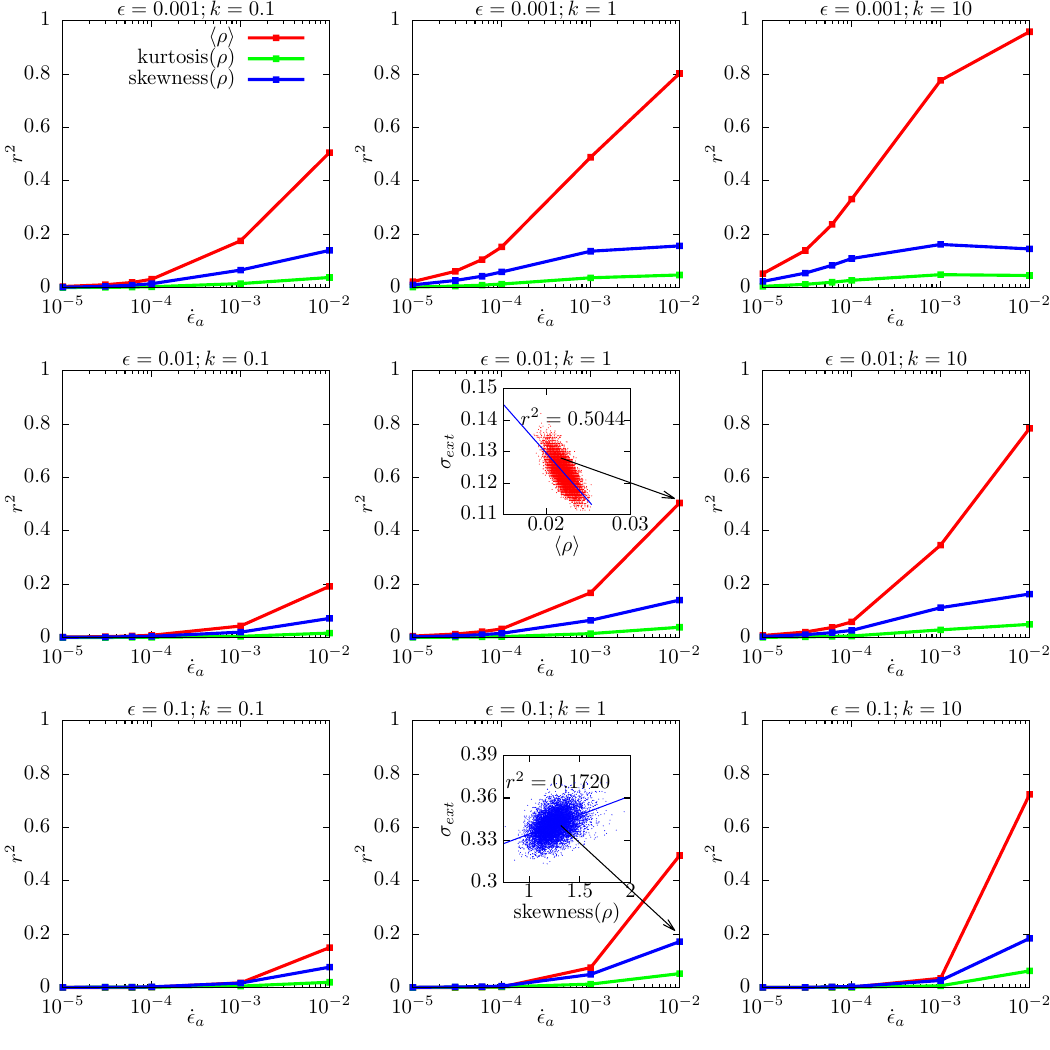}}
	\caption{Coefficients of linear correlation between various properties of the dislocation density field and the values of the external stress obtained at given strain, stiffness and strain rate. Only the features exhibiting significant correlation were shown.}
	\label{linear_correlations_density}
\end{figure}
\begin{figure}[ht]
	\centering
	\resizebox{0.95\columnwidth}{!}{\includegraphics{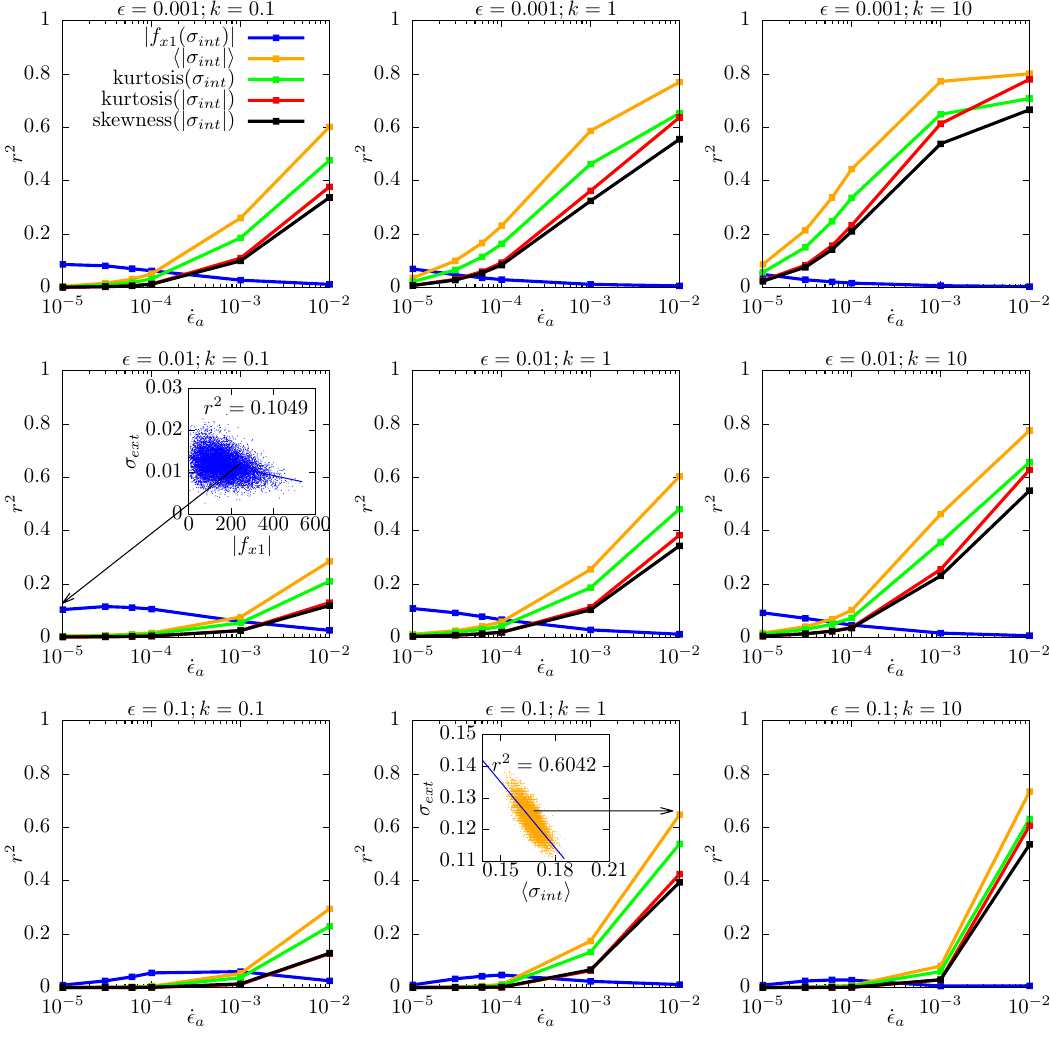}}
	\caption{Coefficients of linear correlation between various properties of the internal stress field and the values of the external stress obtained at given strain, stiffness and strain rate. Only the features exhibiting significant correlation were shown.}
	\label{linear_correlations_stress_field}
\end{figure}
\begin{figure}[ht]
	\centering
	\resizebox{0.95\columnwidth}{!}{\includegraphics{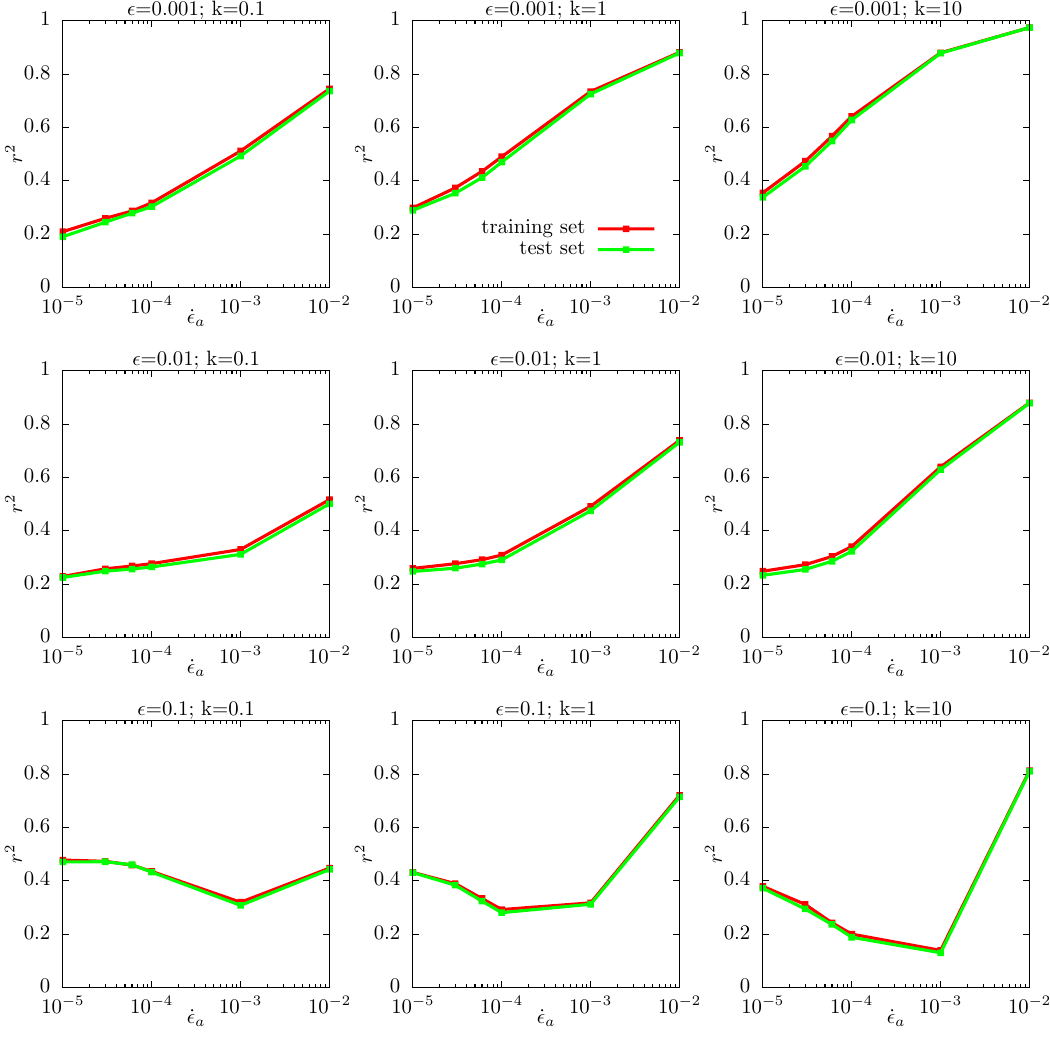}}
	\caption{Scores for the linear regression for the dataset of 10000 configurations averaged over 5 random seeds.}
	\label{LR_scores_10000}
\end{figure}
\begin{figure}[ht]
	\centering
	\resizebox{0.95\columnwidth}{!}{\includegraphics{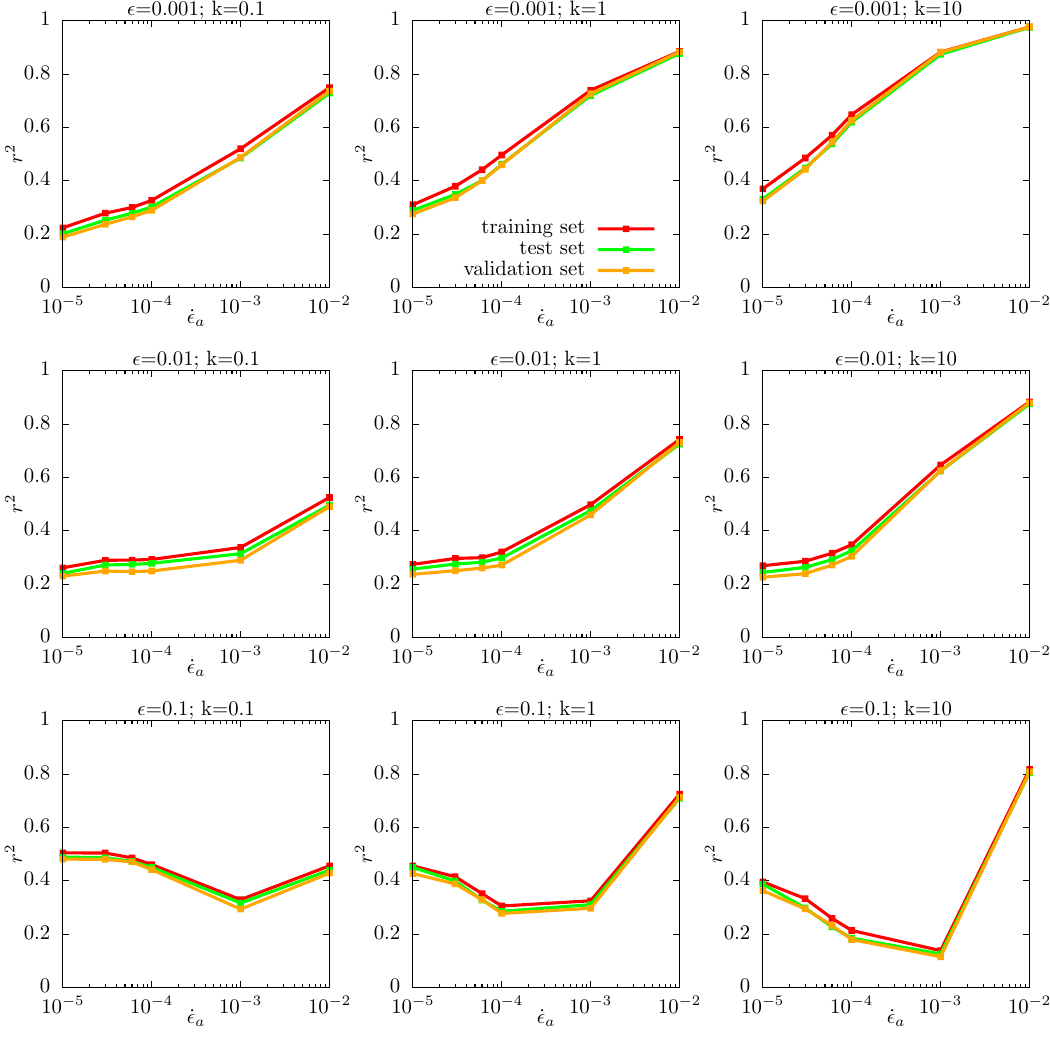}}
	\caption{Scores for the NN for the dataset of 10000 configurations averaged over 5 random seeds.}
	\label{NN_scores_10000}
\end{figure}
\begin{figure}[ht]
	\centering
	\resizebox{0.95\columnwidth}{!}{\includegraphics{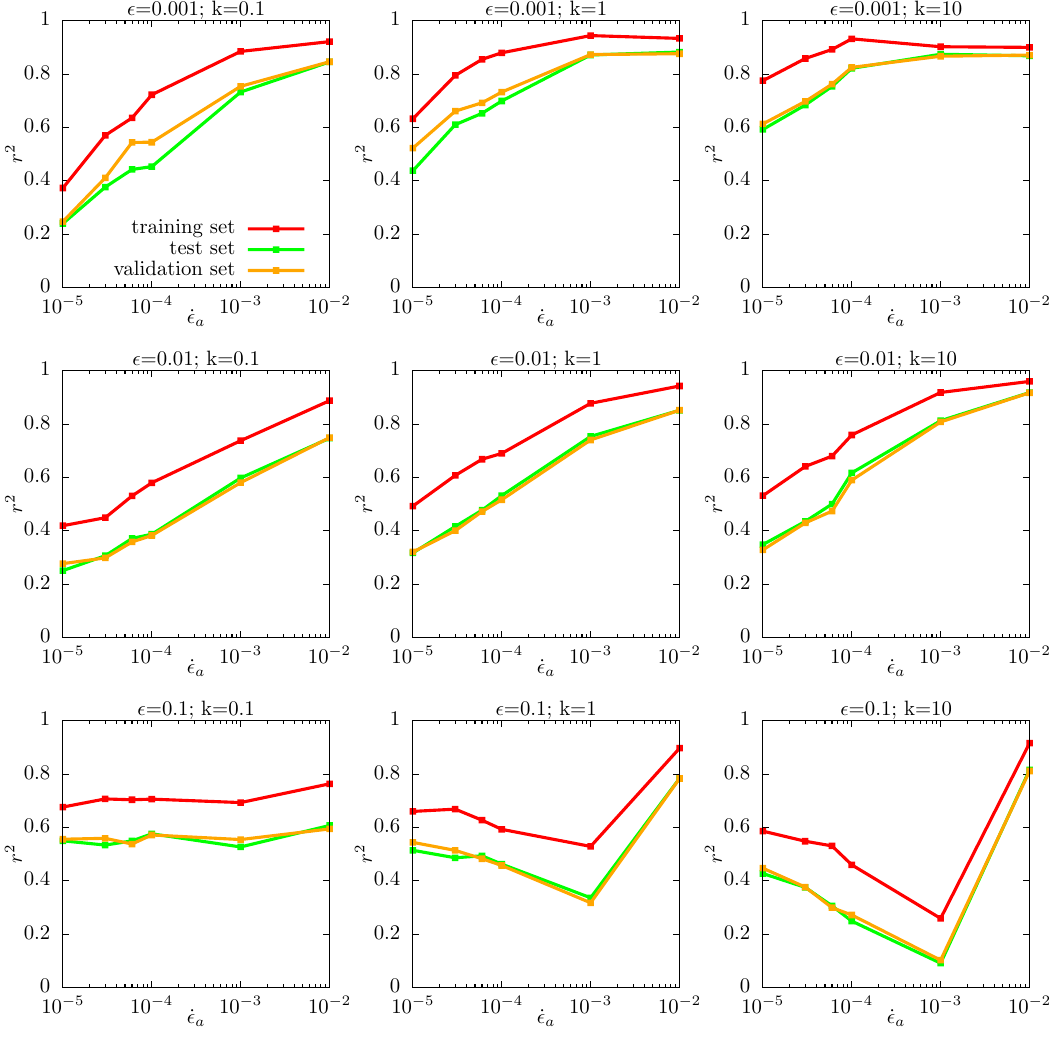}}
	\caption{Scores for the CNN for the dataset of 5000 configurations averaged over 5 random seeds.}
	\label{CNN_scores_5000}
\end{figure}
\begin{figure}[ht]
	\centering
	\resizebox{0.95\columnwidth}{!}{\includegraphics{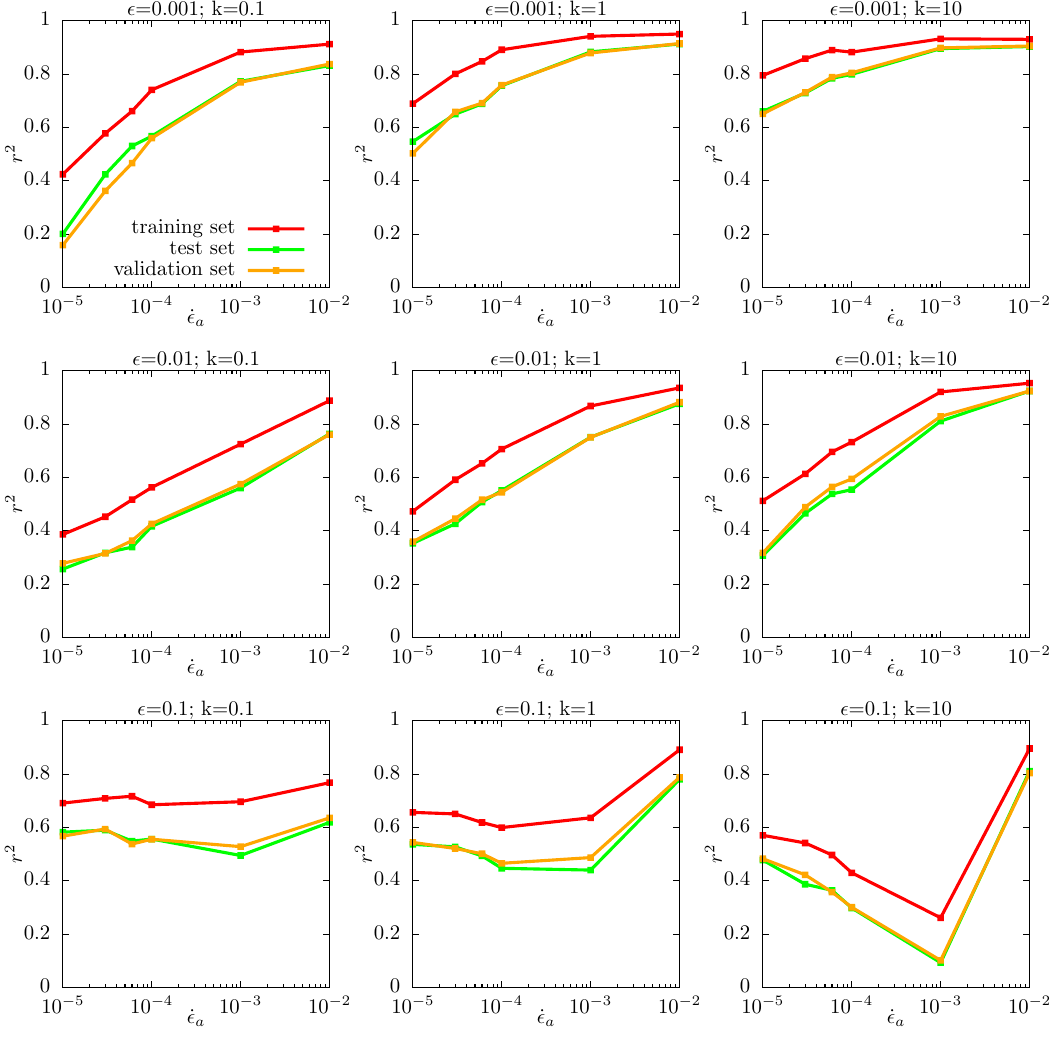}}
	\caption{Scores for the CNN for the dataset of 6000 configurations averaged over 5 random seeds.}
	\label{CNN_scores_6000}
\end{figure}
\begin{figure}[ht]
	\centering
	\resizebox{0.95\columnwidth}{!}{\includegraphics{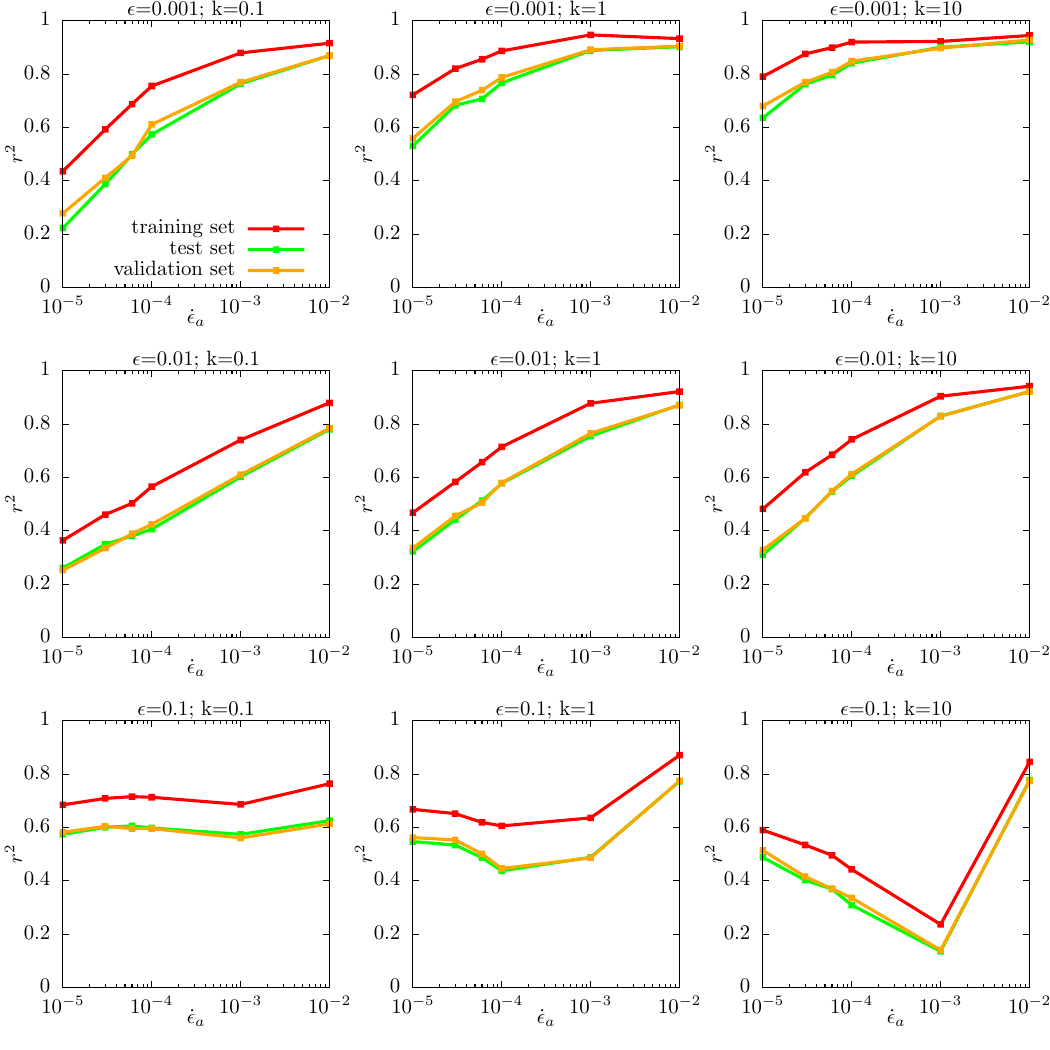}}
	\caption{Scores for the CNN for the dataset of 7000 configurations averaged over 5 random seeds.}
	\label{CNN_scores_7000}
\end{figure}
\begin{figure}[ht]
	\centering
	\resizebox{0.95\columnwidth}{!}{\includegraphics{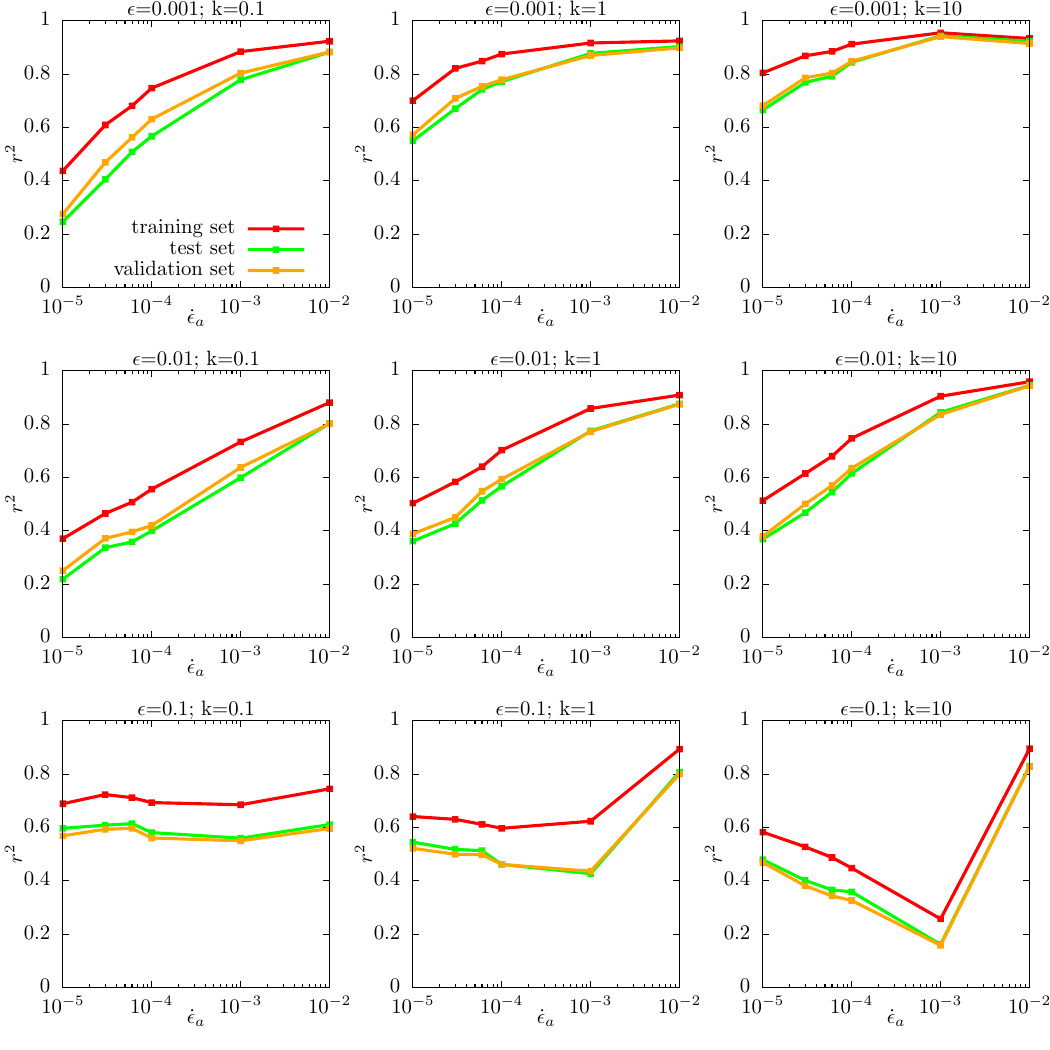}}
	\caption{Scores for the CNN for the dataset of 8000 configurations averaged over 5 random seeds.}
	\label{CNN_scores_8000}
\end{figure}
\begin{figure}[ht]
	\centering
	\resizebox{0.95\columnwidth}{!}{\includegraphics{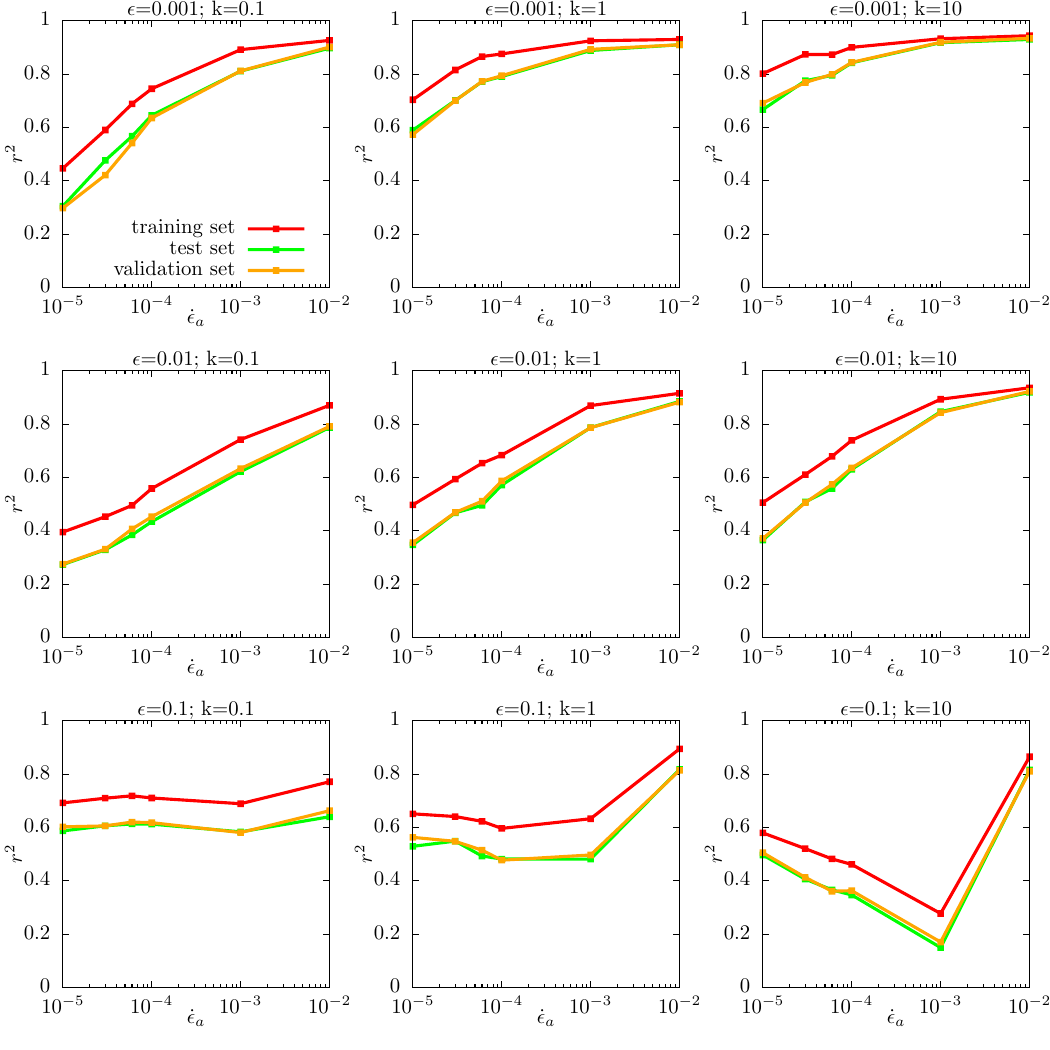}}
	\caption{Scores for the CNN for the dataset of 9000 configurations averaged over 5 random seeds.}
	\label{CNN_scores_9000}
\end{figure}
\begin{figure}[ht]
	\centering
	\resizebox{0.95\columnwidth}{!}{\includegraphics{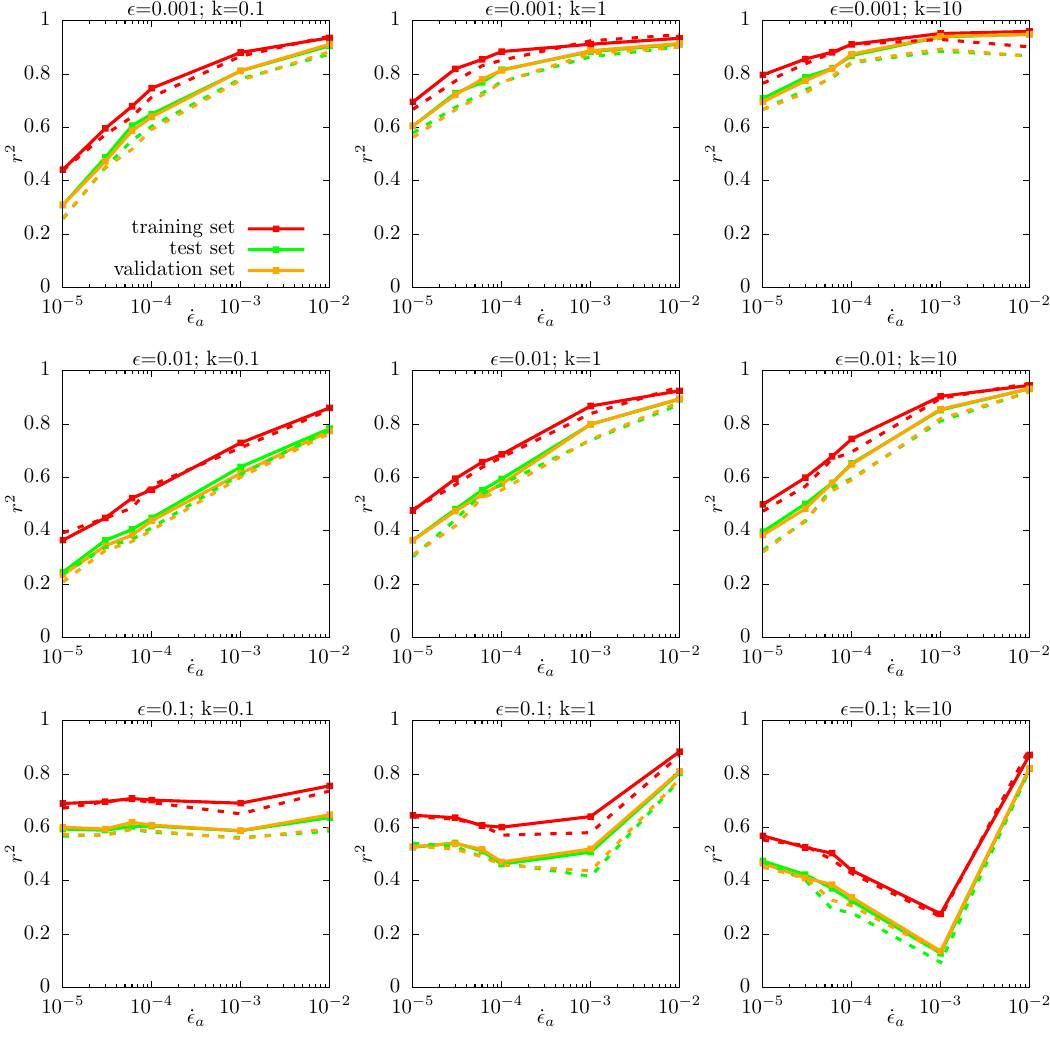}}
	\caption{Scores for the CNN for the dataset of 10000 configurations averaged over 5 random seeds. The dashed lines show the results without L2 regularization compared to those with regularization represented by continuous lines.}
	\label{CNN_scores_10000}
\end{figure}
\begin{figure}[ht]
\centering
	\resizebox{0.95\columnwidth}{!}{\includegraphics{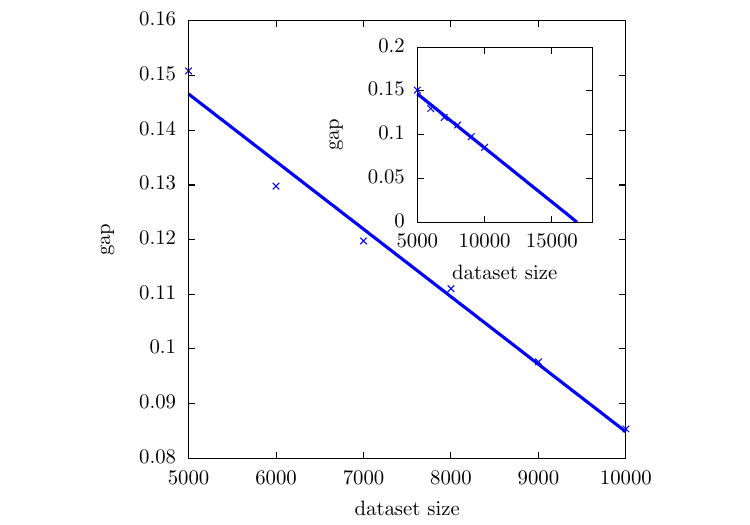}}
	\caption{Gap between the averaged predictability scores of the training and the test set for the CNN as a function of the dataset size. As seen in the inset the extrapolation shows that the gap should completely disappear for the dataset consisting of around 17000 configurations.}
	\label{CNN_gap}
\end{figure}